\documentclass{aa}  

\usepackage[colorlinks=true, linkcolor=blue, citecolor=blue, urlcolor=blue, filecolor=blue]{hyperref} 
\usepackage{graphicx}
\usepackage{txfonts,textcomp}
\usepackage{booktabs}
\usepackage{subcaption}

\defcitealias{J20}{J20}

\begin{document}

   \title{Pre-supernova stellar feedback in nearby starburst dwarf galaxies}
   

   \author{Lucie E. Rowland
          \inst{1,2,3}\fnmsep\thanks{\email{lrowland@strw.leidenuniv.nl}},
          Anna F. McLeod \inst{2,3}, Azadeh Fattahi \inst{2,3}, 
          Francesco Belfiore \inst{4},
          Giovanni Cresci \inst{4},
          Leslie Hunt \inst{4},
          Mark Krumholz \inst{5,6},
          Nimisha Kumari \inst{7},
          Antonino Marasco \inst{8},
          Giacomo Venturi \inst{9,10,4} 
          }

    \institute{Leiden Observatory, Leiden University, P.O. Box 9513, 2300 RA Leiden,
            The Netherlands
        \and
    Centre for Extragalactic Astronomy, Department of Physics, Durham University, South Road, Durham DH1 3LE, UK
        \and
             Institute for Computational Cosmology, Department of Physics, University of Durham, South Road, Durham DH1 3LE, UK
        \and 
            INAF - Osservatorio Astrofisico di Arcetri, Largo E. Fermi 5, I-50125 Firenze, Italy
        \and
            Research School of Astronomy and Astrophysics, Australian National University, Cotter Rd, Weston ACT 2611 Australia
        \and    
            ARC Centre of Excellence for Astronomy in Three Dimensions (ASTRO-3D), Canberra ACT 2600 Australia
        \and
            AURA for European Space Agency (ESA), ESA Office, Space Telescope Science Institute, 3700 San Matin Drive, Baltimore, MD, 21218, USA
        \and 
            INAF - Padova Astronomical Observatory, Vicolo dell’Osservatorio 5, 35122 Padova, Italy
        \and
            Scuola Normale Superiore, Piazza dei Cavalieri 7, I-56126 Pisa, Italy
        \and 
            Instituto de Astrofísica, Facultad de Física, Pontificia Universidad Católica de Chile, Casilla 306, Santiago 22, Chile
             }
   \date{Received ; accepted }


  \abstract
   {Stellar feedback in dwarf galaxies remains, to date, poorly explored, yet is crucial to understanding galaxy evolution in the early Universe. In particular, pre-supernova feedback has recently been found to play a significant role in regulating and disrupting star formation in larger spiral galaxies, but it remains uncertain if it also plays this role in dwarfs.}
   {We study the ionised gas properties and stellar content of individual star-forming regions across three nearby, low-metallicity ($12+\log(\mathrm{O/H}) \sim 7.5$), dwarf ($M_* \sim 40 \times 10^6 M_{\odot}$), starburst ($\log(\mathrm{SFR}) \sim -2.8$) galaxies (J0921, KKH046, and Leo P) to investigate how massive stars influence their surroundings and how this influence changes as a function of environment.}
   {We extracted integrated spectra of 30 HII regions from archival VLT/MUSE integral field spectroscopic observations of these three dwarf starburst galaxies. We fitted the HII regions' main emission lines with Gaussian profiles to derive their oxygen abundances, electron densities, and luminosities, and we used the Stochastically Lighting Up Galaxies (\texttt{SLUG}) code to derive the stellar mass, age, and bolometric luminosity of the stellar populations driving the HII regions. We quantified two pre-supernova stellar feedback mechanisms, namely the direct radiation pressure and photoionisation feedback, and explored how feedback strength varies with HII region properties.}
   {Our findings suggest that stellar feedback has less of an impact on evolved regions, with both the pressure of the ionised gas and the direct radiation pressure decreasing as a function of HII region size (i.e. the evolutionary stage). We find that these stellar feedback mechanisms are also dependent on the metallicity of the HII regions. These findings extend results from stellar feedback studies of more massive star-forming galaxies to the low-mass, low-metallicity regime. In addition, we conclude that the use of stochastic stellar population models significantly affects the relationships found between feedback-related pressure terms and HII region properties, and in particular that non-stochastic models can severely underestimate the bolometric luminosity of low-mass stellar populations.
    }
   {}

    \titlerunning{Pre-supernova stellar feedback in nearby dwarfs}
    \authorrunning{Lucie E. Rowland et al.}
   \maketitle
%

\section{Introduction}
\label{sec:intro}

The effect that stars have on their surroundings is a multi-scale and multi-physics process known as stellar feedback. Massive stars ($\gtrsim$ 8 $M_{\odot}$) are, by far, the dominant sources of stellar feedback (e.g. \citealt{Abbott1982}), emitting vast quantities of ionising photons into the interstellar medium (ISM), imparting momentum via radiation pressure and ultimately ending their lives in supernova (SN) explosions (\citealt{Krumholz2014}). Radiative and mechanical feedback mechanisms from massive stars have been shown to drive the expansion of HII regions (e.g. \citealt{Lopez_2011, Lopez2014,McLeod2019,Mcleod2020,McLeod2021,Olivier_2021}), enrich the ISM (e.g. \citealt{Maiolino2019}), regulate and disrupt star formation (e.g. \citealt{Ostriker2010}; \citealt{chevance2020}), and even govern the dark matter distribution within dwarf galaxies (e.g. \citealt{Maschenko2006, Pontzen2014}). Stellar feedback physics is therefore a central topic in galaxy formation and evolution, and understanding it is a critical step towards creating realistic simulations from the scales of individual star-forming regions and giant molecular clouds (GMCs) up to the scales of entire galaxies (e.g. \citealt{Kim_2018,Dale_2014,Hopkins_2018}). 

Recently, focus has shifted from the effects of SNe, which are generally thought to be the dominant stellar feedback mechanism in shaping galaxies on large scales, to so-called pre-SN mechanisms. These mechanisms have been shown to play a significant role in the regulation of star formation prior to the first SN events (\citealt{Lucas2020}; \citealt{McLeod2021}; \citealt{Chevance2022}). Whilst significant progress has been made in recent years in the incorporation of these stellar feedback mechanisms into simulations (e.g. FIRE, \citealt{Hopkins2022}), we currently lack a detailed understanding of their dependence on the physical properties of the ISM. This is particularly the case for dwarf galaxies, which are largely absent from recent large extragalactic surveys (e.g. SAMI, \citealt{Bryant2015}; CALIFA, \citealt{S_nchez_2012}; PHANGS, \citealt{Emsellem2022}; and \citealt{Barnes2021}). 

Dwarf galaxies are interesting targets for feedback studies for a number of reasons. Firstly, they are the most abundant type of galaxy in the Universe (e.g. \citealt{Phillipps1998}), and yet their formation and evolution is still not completely understood (\citealt{Tolstoy_2009}). Secondly, at the low-mass regime of dwarfs, it is well known that the effects of feedback processes are more dramatic due to galaxies' shallow gravitational potentials  (e.g. \citealt{Hopkins_2014}). Finally, based on the mass-metallicity relation (e.g. {\citealt{Lequeux_1979}}), we can also expect the HII regions within a sample of low-mass galaxies to be metal-poor, providing a comparison sample to the more metal-rich HII regions in recent works (e.g. \citealt{Barnes2021} and \citealt{DellaBruna2022}) and allowing us to extend the study of stellar feedback effects across a wider range of environments. This makes nearby dwarf galaxies critical laboratories for providing observational constraints necessary for more realistic simulations.

In the past decade or so, observational studies of stellar feedback have been made increasingly more detailed by the development of integral field unit (IFU) spectroscopy. IFU instruments are ideal to study the gaseous and stellar components of resolved star-forming regions, allowing us to derive the properties and kinematics of the feedback-affected ionised gas, to derive the properties of the feedback-driving stellar populations, and to quantify the feedback mechanisms themselves. This use of IFU spectroscopy has been demonstrated in recent works, including \citet{Lopez2014}, \citet{McLeod2021}, \citet{Barnes2021}, and \citet{DellaBruna2022}. This ever-increasing number of spatially resolved studies of star-forming regions within our own Galaxy and in nearby star-forming galaxies has created a vast database of HII regions, which in this paper we extend to low-mass, low-metallicity host galaxies.

Consequently, in this work we aim to investigate a sample of nearby, low-metallicity dwarf starburst galaxies observed with the Multi Unit Spectroscopic Explorer (MUSE, \citealt{bacon2010}) IFU instrument mounted on the Very Large Telescope (VLT). We then make comparisons to similar studies in more massive systems, for example, the sample of more massive galaxies from the Physics at High Angular resolutions in Nearby GalaxieS (PHANGS) MUSE survey (\citealt{Emsellem2022} and the corresponding stellar feedback studies of \citealt{Barnes2021, Barnes2022}), in order to contribute to the ongoing investigation of the environmental dependence of stellar feedback.

This paper is organised as follows. We give a brief overview of our selection of dwarf galaxies and of the VLT/MUSE observations in Section \ref{sec:obs}, followed by Section \ref{sec:data analysis} where we describe the techniques used to select, classify, and extract the integrated spectra from our sample of 30 HII regions. In Section \ref{sec:ion_gas}, we describe the methods used to derive the ionised gas properties of these regions. In Section \ref{sec:SPS}, we discuss the stellar population synthesis modelling, and how we used these libraries of simulations to derive star cluster properties within our regions. We then explain how we derived two pre-SN stellar feedback mechanisms in Section \ref{sec:quantifying}, and discuss  our findings and compare them to those of other works in Section \ref{sec:Discussion} in order to quantify these mechanisms as a function of environmental properties. Finally, in Section \ref{sec:conc} we give a summary of our work and conclusions.

\begin{figure*}[t]
    \centering
    \small
    \includegraphics[angle=270,width=\textwidth]{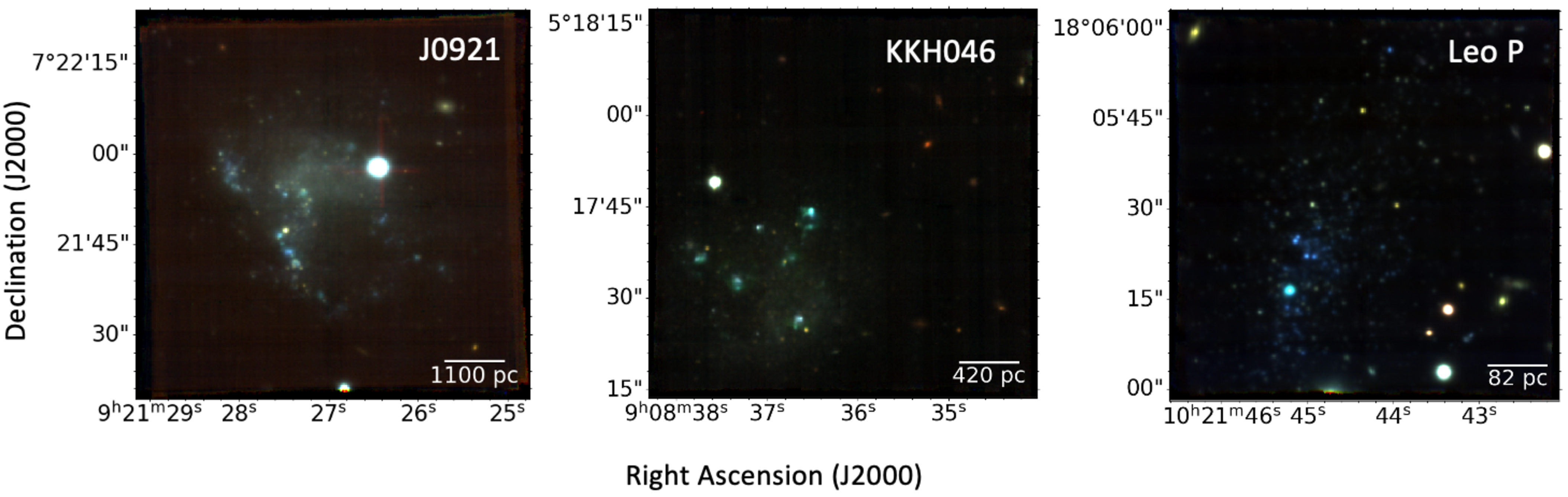}
    \caption {Three-colour composite images of the three dwarf galaxies (left: J0921, centre: KKH046, right: Leo P) collapsed across three arbitrary wavelength ranges and coloured blue (4650-5800 \AA), green (5800-7000 \AA), and red (7000-8200 \AA). }
    \label{fig:galaxies}

\end{figure*}

\begin{figure*}[t!]
    \centering
    \includegraphics[width=\textwidth]{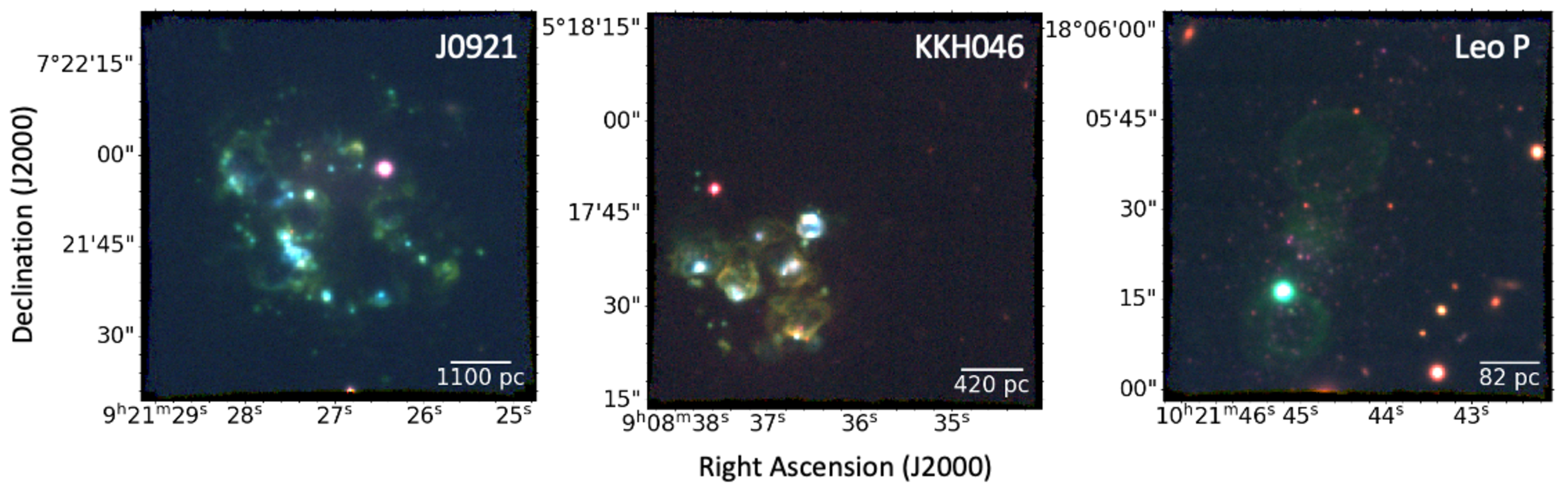}
    \small
    \caption {Same as Figure \ref{fig:galaxies}, except here red corresponds to [SII]$\lambda$6716, green to H$\alpha$, and blue to [OIII]$\lambda$5007 emission. }
    \label{fig:ionised gas maps}
\end{figure*}

\section{Observations and target selection}
\label{sec:obs}

The dwarf galaxies analysed in this paper were selected from the DWarf galaxies Archival Local survey for Ism iNvestigation (DWALIN, \citealt{Marasco2022}, Cresci et al. in prep) galaxy sample. DWALIN consists of 40 nearby (median distance $D \sim$ 9.3 Mpc), low stellar mass (median $M_* \sim 10^8 ~\mathrm{M}_{\odot}$), highly star-forming (median specific star formation rate, $\mathrm{sSFR} \sim 10^{-9.44} \mathrm{yr}^{-1}$) galaxies with archival VLT/MUSE data. MUSE is an optical IFU, with a wavelength range of 4600-9350 $\mathrm{\AA}$ and a 1'${\times}$1' field-of-view in its extended wide field mode (WFM), a pixel scale of 0.2" pixel $^{-1}$, and a resolving power of ${\sim}$ 1770-3590. As described in \citet{Marasco2022}, the data reduction for the DWALIN sample was carried out via the MUSE pipeline (\citealt{Weilbacher2020}), v2.8.1. Full details of the sample and data reduction will be given by Cresci et al. (in prep). 

From the DWALIN sample, we have selected three dwarf galaxies, namely AGC 193816 (also known as J0921+0721 and henceforth referred to as J0921), KKH046 and Leo P. These three galaxies show clearly visible ionised gas bubbles, which are the signatures of feedback and hence suitable for this study. Observations were taken under programmes 096.B-0212 (PI James), 0104.D-0199 (PI Brinchmann), and 094.D-0346 (PI Evans) for J0921, KKH046 and Leo P, respectively. In Figures \ref{fig:galaxies} and \ref{fig:ionised gas maps}, we show three-colour composite images of these galaxies. The images in Figure \ref{fig:galaxies}, which trace the stellar populations of these galaxies, are produced by dividing the wavelength range of MUSE intro three arbitrary colour ranges (blue = 4650-5800 \AA, green = 5800-7000 \AA, red = 7000-8200 \AA) and integrating the cubes across these three ranges to produce and combine the images in each custom filter. We exclude the red channels at 8200-9350 \AA ~and the blue edge channels at 4600-4650 \AA, where there are some residual background effects, from our analysis. In Figure \ref{fig:ionised gas maps}, we combine three different emission line maps (red = [SII]$\lambda$6717, green = H$\alpha$, blue = [OIII]$\lambda$5007) to trace the ionised gas within these galaxies. These emission maps are obtained by integrating the MUSE cube over a 6 \AA ~wide spectral window centred on the corresponding emission line. The H$\alpha$ emission maps are also shown in Figure \ref{fig:H alpha maps}. We selected these three galaxies from the DWALIN sample on the basis that they are observed with $< 100$ pc resolution, and the importance of this is apparent from the emission maps, which at this resolution show clearly distinct bubbles of ionised gas.

The gas-phase metallicities of these galaxies, as measured by their oxygen abundance ($12+ \log(\mathrm{O/H})$), are given in Table \ref{tab:galaxies}, along with other key properties. We also give the spatial resolutions, $\Theta$, of the VLT/MUSE observations used in this work as determined from the distance to each galaxy, $D$, and the reported seeing of the observations. We note that the seeing for the observations of KKH046 were considerably improved by the use of the VLT/MUSE Adaptive Optics (AO) system. As in \citet{Marasco2022}, the distances for KKH046 and J0921 are taken from \citet{Kourkchi20} (where distances are determined using the local peculiar velocity model of \citealt{Masters2005}) and for Leo P we use the value determined by \citet{McQuinn_2015} for reasons described therein.

\begin{table*}[t]
\caption[]{Properties of the target dwarf galaxies in this paper.}
\normalsize
\centering
\begin{tabular}{lllllllll}
\toprule

\centering

   Galaxy & RA, Dec & $D$   &  log($\frac{SFR}{M_{\odot}\mathrm{yr}^{-1}}$) & 12+log(O/H) & log($\frac{\mathrm{M_{HI}}}{\textup{M}_\odot}$)& log($\frac{\mathrm{M_{\star}}}{\textup{M}_\odot}$)  & $\Theta$  \\
    & (J2000) & (Mpc) & & & & & (pc)\\
        \hline\hline
        J0921  &  09:21:27.2, +7:21:50.7 &   $21 \pm 4 ^a$ & -1.64 $\pm$ 0.10 $^b$ &  $7.67 \pm 0.20 ^c$  & $8.44 \pm 0.10 ^d$ & $8.03 \pm 0.22 ^b$ &83 \\
        \hline
        KKH046 & 09:08:36.2, +5:17:45.5 & $12 \pm 2 ^a$ & -2.37 $\pm$ 0.10 $^b$ & $7.68 \pm 0.03 ^e$  & $7.74 \pm 0.23 ^a$ & $7.09 \pm 0.20 ^b$ & 32 \\
        \hline
        Leo P  & 10:21:45.1, +18:05:17.0 &  $1.6 \pm 0.1 ^f$ & -4.40 $\pm$ 0.38 $^b$ & $7.17 \pm 0.04 ^f$  & $5.96 \pm 0.10 ^f$ & $5.75 \pm 0.30 ^f$ & 9.8\\
        
\bottomrule
\end{tabular}

\label{tab:galaxies}
\tablefoot{
    Columns are (2) their J2000 coordinates, (3) the distance to the galaxy in megaparsecs, (4) their star formation rate (SFR), (5) their oxygen abundance as derived from the direct method, (6) the log of their HI mass in solar mass units, (7) the log of their stellar mass in solar mass units, and (8) the spatial resolution of the MUSE observations. Values are taken from $^a$ \citet{Kourkchi20}, $^b$ \citet{Marasco2022}, $^c$ \citet{James2015},$^d$ \citet{Haynes2018}, $^e$ \citet{Izotov2012}, and $^f$ \citet{McQuinn_2015}.}
\end{table*}

\section{Data analysis}
\label{sec:data analysis}
\subsection{HII region selection}
\label{sec:hii_regions}

\begin{figure*}[t!]
    \centering
    \small
    \includegraphics[angle=270,width=\textwidth]{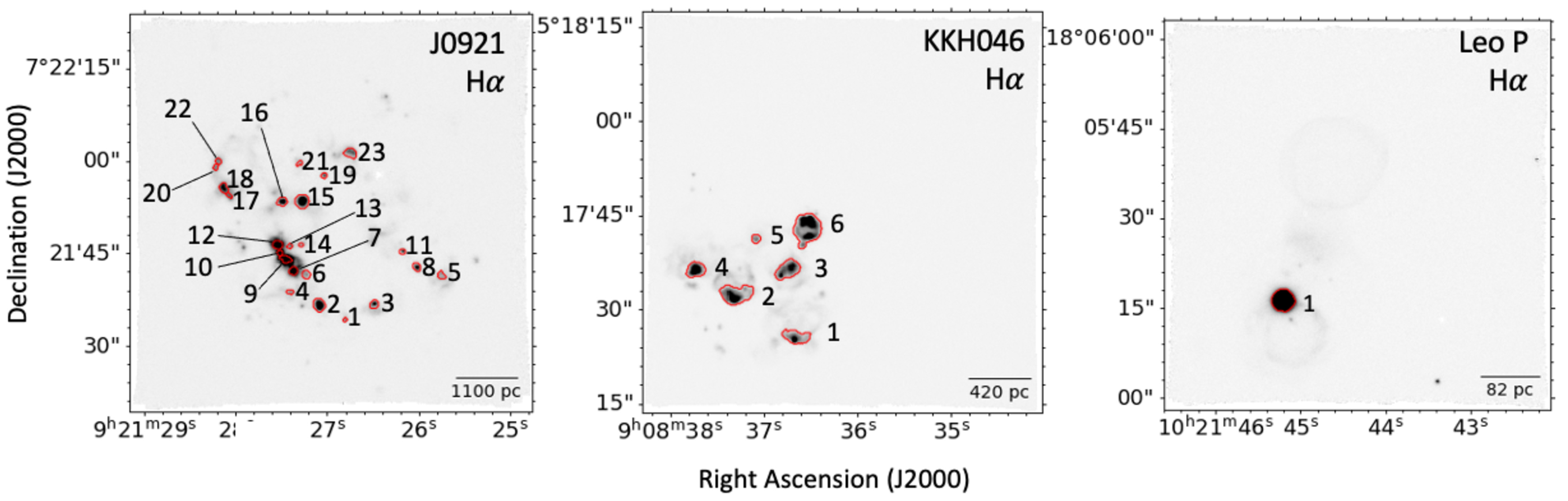}
    \caption {H$\alpha$ emission maps of J0921 (left), KKH046 (centre), and Leo P (right), with the dendrogram structures identified in this work as HII regions outlined in red. Each HII region is labelled by a number on its right or via a line.}
    \label{fig:H alpha maps}
\end{figure*}

\begin{figure*}[h]
    \centering

    \begin{subfigure}{0.48\textwidth}
        \includegraphics[width=0.6\textwidth,angle=270]{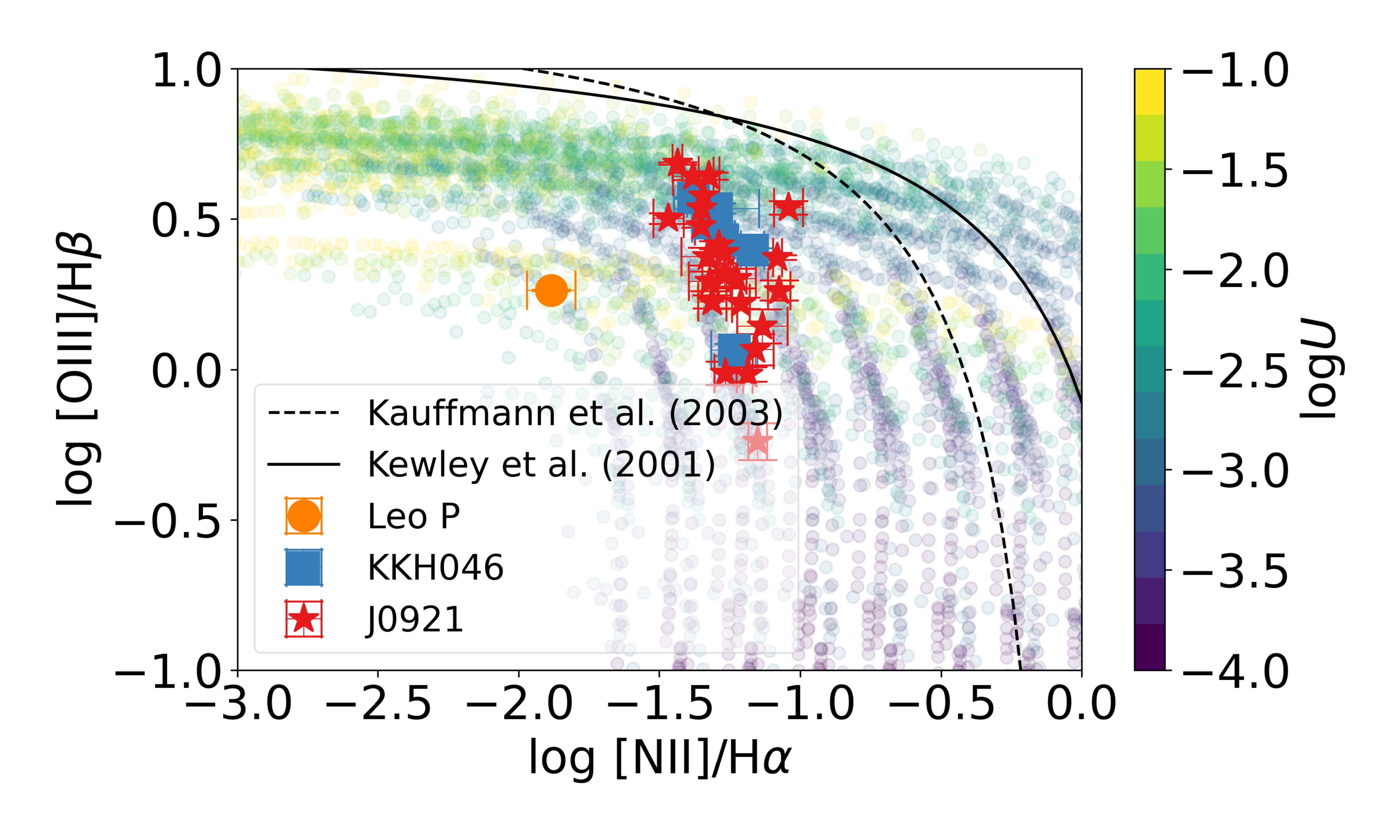}
    \end{subfigure}
    \begin{subfigure}{0.48\textwidth}
        \includegraphics[width=0.6\textwidth,angle=270]{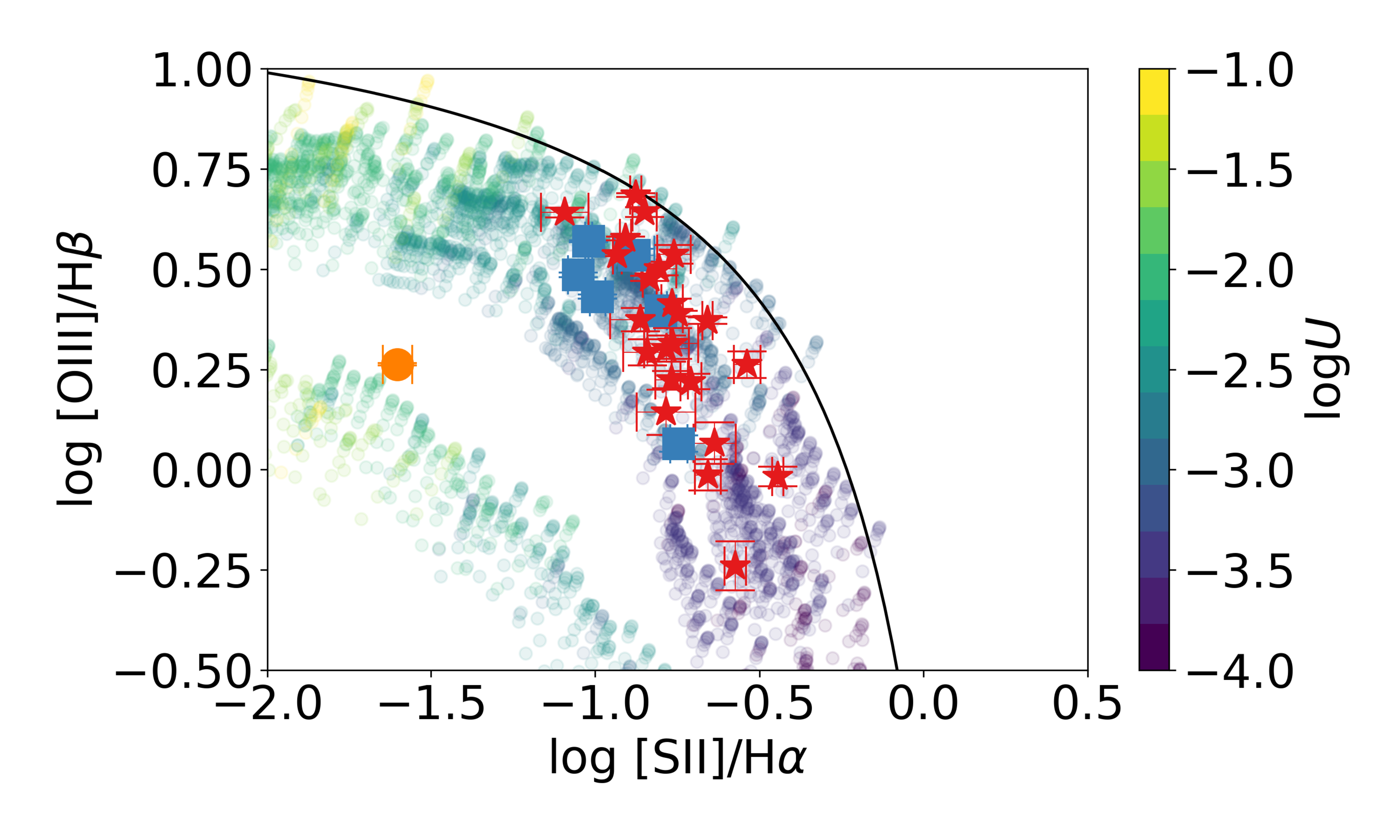}
    \end{subfigure}
    \caption {Classical emission line diagnostic diagrams of the HII regions defined in this paper. Theoretical maximum starburst lines plotted by the solid and dashed black lines are taken from literature (\citealt{Kewley2001} and \citealt{Kauffmann2003}, respectively). Low-metallicity photoionisation models from \texttt{CLOUDY} are also plotted, coloured by ionisation parameter ($\log U$).}
    \label{fig:BPT}
\end{figure*}

\begin{figure*}[h]
    \centering
    \includegraphics[width=0.33\textwidth,angle=270]{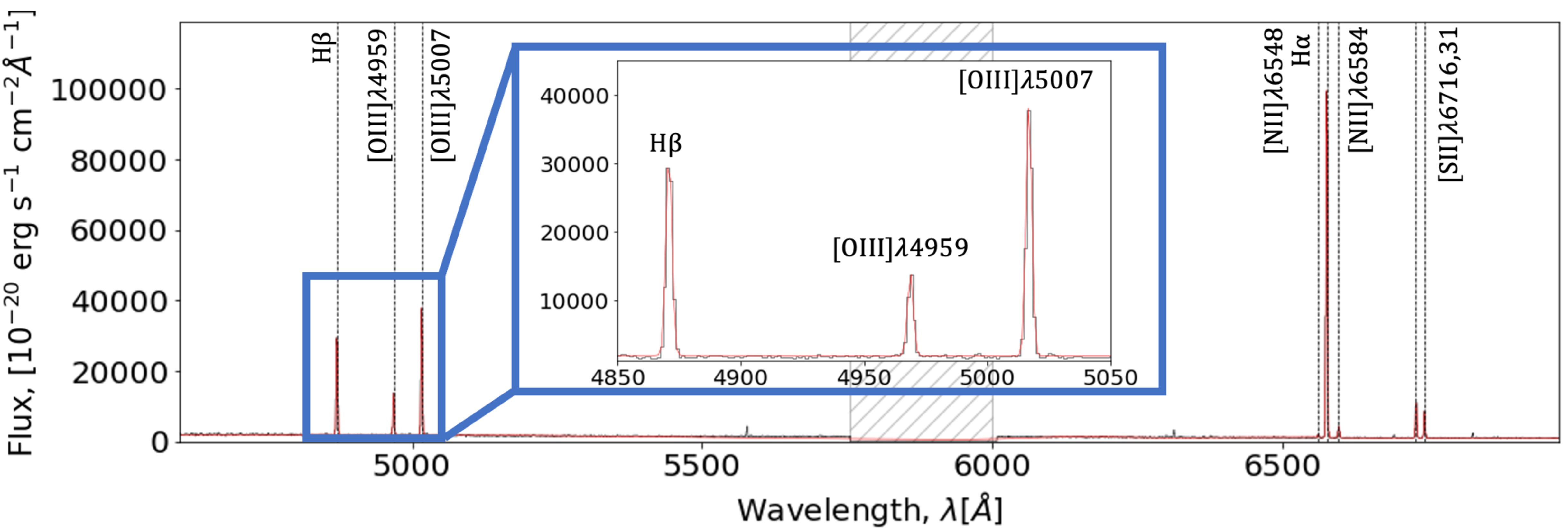}
    \caption{Portion of the MUSE spectrum of HII region 1 in KKH046 which covers the  key nebular emission lines studied in this paper, with the \texttt{PYSPECKIT} Gaussian fit plotted in red. The hashed region indicates the AO notch filter. In the overlaying panel outlined in blue, we show a closer look at the H$\beta$ and [OIII]$_{\lambda4959,5007}$ doublet lines.}
    \label{fig:KKH046 spectrum}
\end{figure*}

As in \citet{McLeod2021}, we use H$\alpha$ emission line maps with the \texttt{ASTRODENDRO}\footnote[1]{\url{http://www.dendrograms.org/}} Python package to trace the boundaries of HII regions within the target galaxies and extract their spatially integrated spectra. These emission maps are continuum corrected by subtracting a continuum map, obtained by integrating a small section of the continuum level either side of the H$\alpha$ line with no other significant nebular emission or sky lines. The \texttt{ASTRODENDRO} package then takes these 2D emission maps and returns tree-like hierarchical structures called dendrograms, with areas of bright emission called `leaves'. These structures are strongly dependent on user-defined parameters, namely the minimum flux threshold necessary for a pixel to be added to a structure, $F_{min}$, and the minimum number of pixels contained within an independent structure, $N_{min}$. 

As can be seen in Figures \ref{fig:ionised gas maps} and \ref{fig:H alpha maps}, there are some clear bright clumps of ionised gas in each galaxy, along with fainter diffuse ionised gas (DIG) emission surrounding these clumps. This surrounding DIG can represent a significant fraction of the total H$\alpha$ emission (\citealt{Oey2007}, \citealt{Kumari2020}). For example, in their recent analysis of MUSE data of M83, \citet{Della_Bruna_2021} find a DIG fraction of $\sim 27-42 \%$. In comparison to HII regions, DIG is found to have a lower surface brightness, lower electron density ($\sim$ 0.1 cm$^{-3}$) and slightly lower ionisation states (e.g. \citealt{Mannucci2021}; \citealt{Belfiore2022} and references therein). The origin of the DIG found in the ISM of galaxies is still not fully understood, though suggested explanations include the leakage of photons from HII regions (e.g. \citealt{Weilbacher2018}), ionisation by shocks (e.g. \citealt{Dopita1995}), and ionisation by evolved, post AGB stars (e.g. \citealt{Byler2019}), or a combination of the above (\citealt{FloresFajardo2011, Belfiore2022}). The DIG can therefore show line ratios that differ significantly from those seen in HII regions, and contamination from DIG can substantially impact the physical properties derived from these line ratios for both HII regions, and for galaxies as a whole (\citealt{Zhang2017}). To limit this contamination, we therefore only consider regions with significant H$\alpha$ emission and of reasonable physical size (exact thresholds described below).

For KKH046 and Leo P, we set the minimum number of pixels within a region ($N_{min}$) to be comparable to the size of the seeing disc for the reported seeing of each observation, and we set the minimum flux threshold ($F_{min}$) to twice the standard deviation of the H$\alpha$ flux within each galaxy. These $F_{min}$ values correspond to emission measures of $\sim 120$ pc cm$^{-6}$, which are greater than the cutoff values used in \citet{McLeod2021} for NGC~300 ($\sim 60$ pc cm$^{-6}$) and in \citet{Della_Bruna_2021} for NGC~7793 ($\sim 82$ pc cm$^{-6}$). However, a test of lower thresholds for KKH046 produced large, irregularly shaped structures and multiple small leaves found to have negative electron densities, which could indicate potential contamination from DIG. We find that the criteria stated above produce regions of reasonable size around each visibly distinguishable ionised gas bubble within KKH046 and Leo P, that is, greater than the spatial resolution of the observations and comparable to similar studies of extragalactic HII regions. For example, \citet{Barnes2021} find a range in HII region radii of 16.6-388.2 pc in PHANGS-MUSE galaxies, which is comparable to the sizes obtained for HII regions in this study (22-270 pc). 

The emission maps of J0921 reveal a more complex structure of multiple scattered HII regions that are less well defined by eye, likely due to it being the most distant galaxy selected. For an initial estimate of the number of regions within this galaxy, we consider the analysis of J0921 carried out by \citet{J20} (hereafter referred to as \citetalias{J20}), who found 30 HII regions with the same MUSE data. A discussion of this paper with a comparison to some of our results is given in Appendix \ref{appendix:j0921 comparison}. With an input of $F_{min}$ of three standard deviations above the median H$\alpha$ flux, or 6.90 $\times 10^{-18}$ erg s$^{-1}$ cm$^{-2}$, and an $N_{min}$ of 6 pixels, we find 25 HII regions, comparable to those found by J20. This flux threshold corresponds to an emission measure of 82 cm pc$^{-6}$, that is, comparable to values used in \citet{McLeod2021} and \citet{Della_Bruna_2021} for NGC~300 and NGC~7793, respectively. However, as we progressed through our analysis, we found that four of these J0921 regions seemingly had unphysical, negative electron densities. This could be for a number of reasons, including contamination from DIG. For two of these regions, numbered 6 and 20 in Figure \ref{fig:H alpha maps}, we found that the dendrogram regions were too large compared to their discernible size, and we therefore use smaller, by eye, circular apertures. We discard the other two regions from subsequent calculations, although further investigation into their sources of ionisation would be interesting for future study. This results in 23 HII regions for J0921, though we note that four regions are smaller than the spatial resolution of the observations, increasing the uncertainty in the properties we have derived.

To test our confidence in the sizes of the dendrogram leaves, we calculated the radius that encompasses 90\% of the H$\alpha$ flux ($R_{90}$) of each visible clump; an example of this process is given Appendix \ref{appendix:radial intensity profiles}. The $R_{90}$ of each clump are found to be comparable to the radius of our final dendrogram structures (measured by the geometric mean of the major and minor axes), which are listed in column 2 of Table \ref{tab:results}. We note that we do not apply a PSF correction due to the lack of high resolution imaging available for all three galaxies, but the observations are seeing-limited and at the spatial resolution of J0921, these HII regions are more likely to be unresolved. 

In order to confirm our classification of these dendrogram leaves as HII regions, rather than supernova remnants (SNRs) or planetary nebulae (PNe), we investigate their location in  Baldwin, Phillips and Terlevich (BPT) diagrams (\citealt{Baldwin1981}). These diagnostic plots were initially introduced to separate star-forming galaxies from objects excited by harder ionising sources (e.g. active galactic nuclei, AGNs), but are also useful to investigate the ionisation state and nature of individual regions of ionised gas within galaxies (e.g. \citealt{espinosa2020}; \citealt{McLeod2021}; \citealt{Belfiore2022}). SNRs typically show higher [NII]/H$\alpha$ ratios, while PNe tend to show enhanced [OIII]/H$\beta$ ratios (e.g. \citealt{Baldwin1981}). We do not find evidence of these enhanced ratios, and all our regions are found to lie below the theoretical `starburst' lines from \citet{Kewley2001} and \citet{Kauffmann2003} (Figure \ref{fig:BPT}), indicating that the dominant source of ionisation within these regions is photoionisation, as expected for HII regions. 

However, since the interpretation  of these diagrams is dependent on a number of factors (see \citealt{Kewley2019b}) and due to the low metallicity of these systems, further investigation with photoionisation modelling is necessary to more confidently infer their ionisation states (\citealt{Plat_2019}). We have made some comparisons to low-metallicity \texttt{CLOUDY} (\citealt{CLOUDY2017}) photoionisation models from the Mexican Million Models database (\citealt{3mdb}), shown by the small circular markers in Figure \ref{fig:BPT} coloured in terms of their ionisation parameter, log$U$. We choose to plot only models with HII regions aged 6 Myr, since this is closest to the median age found for our clusters (see Section \ref{sec:SPS}). We plot models for 12+log(O/H) ranging from 7.0  to 8.0, although we note that the models in this database have a gap between 7.0-7.4, which is the metallicity range that covers Leo P. In general, however, we find that the flux ratios of the regions studied in this work appear to be compatible with low-metallicity photoionisation models, supporting their classification as HII regions.

Overall, our selection strategy produces 23 HII regions within J0921, six within KKH046, and one in Leo P, with sizes ranging from 22 to 270 pc. These HII regions are outlined by red contours and numbered in Figure \ref{fig:H alpha maps}.

We note that when making comparisons between these three galaxies, and with HII regions in the literature, the change in the spatial resolution of the observations may introduce some bias to the interpretations. For example, the derived radii of HII regions in nearby targets, such as the Large Magellanic Cloud (LMC) and Small Magellanic Cloud (SMC), or within the Milky Way, are more likely to appear smaller than regions in more distant targets due to the ability to resolve finer detail. This study would therefore benefit from higher spatial resolution follow-up observations to reduce the uncertainty on the sizes of these regions and to reveal more details of their morphology, for example, as carried out with Hubble Space Telescope (HST) high-resolution observations in \citet{Barnes2022}. We also introduce a bias by selecting only the brightest H$\alpha$ emission regions, as we could be missing weak HII regions. These biases must be especially considered in Section \ref{sec:Discussion}, where we make comparisons to studies with smaller and more compact HII regions.

\subsection{Emission line fitting}
\label{sec:line_fitting}

We next use the \texttt{SpectralCube} package (\citealt{SpectralCube}) to extract the spectra of the individual regions within the dendrogram leaf contours shown in Figure \ref{fig:H alpha maps}. We continuum-correct the spectra by subtracting a linear baseline fit using the \texttt{PYSPECKIT} Python package (\citealt{Ginsburg2011}). Figure \ref{fig:KKH046 spectrum} shows a portion of the MUSE spectrum for region 1 in KKH046, as an example. We note that the hashed rectangle in the spectrum indicates a gap in the spectral coverage due to the notch filter from the AO system used for the observations of KKH046. We fit the key nebular lines (e.g. H$\alpha$, H$\beta$, [OIII]$\lambda$5007,4959, [NII]$\lambda$6548,6584, and [SII]$\lambda$6716,6731) with Gaussians using \texttt{PYSPECKIT}, and we show an example fit to the [OIII] doublet and H$\beta$ line in Figure \ref{fig:KKH046 spectrum}. To correct for attenuation, we assume an attenuation curve for the continuum of starburst galaxies empirically derived in \citet{Calzetti2000}, and calculate the colour excess, $E(B-V)$, from the Balmer decrement of each HII region assuming Case B recombination with an intrinsic ratio of 2.86 (\citealt{Osterborck2006}). \citet{Calzetti1994} showed that HII regions are more heavily attenuated than the stellar continuum, and so we adopt their derived conversion factor of 2.27 between the stellar continuum attenuation curve and the nebular attenuation. The continuum-subtracted, attenuation-corrected fluxes are then used for all following calculations, with the uncertainties returned by \texttt{PYSPECKIT} propagated through. We also determine the visual extinction, $A_V$, for each of our HII regions from the product of the colour excess and the total-to-selective attenuation, for which we adopt the value of 4.1$\pm$0.8 derived by \citet{Calzetti2000} for nearby starbursts. We note that the results presented in this paper do not significantly change when a Milky Way extinction curve (\citealt{Cardelli1989}) or LMC extinction curve (\citealt{Fitzpatrick1999}) are used.

\section{Ionised gas properties}
\label{sec:ion_gas}
In this section, we focus on the methods used to derive the electron densities, luminosities, and oxygen abundances of the HII regions in our sample. These properties, along with the region size, extinction, and stellar content, are of particular interest for the purpose of studying feedback as a function of environment. We list some of the key ionised gas properties, the feedback-related pressure terms, and the stellar population properties derived in this work in Table \ref{tab:results}, along with their estimated uncertainties. For a more in-depth discussion of our estimated uncertainties, see Appendix \ref{appendix:uncertainties}. 

\subsection{Electron densities}

\label{sec:e_dens}
We derive the electron density ($n_e$) from the line ratio $R_{[SII]}$ = [SII]$_{\lambda 6716}$/[SII]$_{\lambda 6731}$ using the calibration of \citet{McCall1984}, given by 
\begin{equation}\label{eq:density}
    n_e = \frac{R_{[SII]}-1.49}{5.6713-12.8 R_{[SII]}}.
\end{equation}

\noindent This assumes an electron temperature of $10^4$ K, which is typical of nearby HII regions (e.g. \citealt{Peimbert_2017}), including some studies of low-metallicity HII regions (e.g. \citealt{J20}). As listed in Table \ref{tab:results}, we find an electron density range of $\sim$ 5-260 cm$^{-3}$. \citet{Kewley2019ac} find that the use of $R_{[SII]}$ as an electron density diagnostic is most effective within the range $\sim$ 40-10$^4$ cm$^{-3}$. For the HII regions with densities below this lower threshold, these values are likely to be more uncertain.

\subsection{Luminosities}
\label{sec:lum}

We derive the H$\alpha$ luminosities, $L_{\mathrm{H\alpha}}$, from the total integrated H$\alpha$ flux of each HII region and assuming distances as given in Table \ref{tab:galaxies}. In order to derive the direct radiation pressure, $P_{\mathrm{dir}}$ (see Section \ref{sec:Pdir}), for each HII region, we are also interested in deriving a value for the bolometric luminosity, $L_{\mathrm{bol}}$, of the stellar population within each region. \citet{Kennicutt2012} defines a method of estimating the bolometric luminosity from the observed, extinction-corrected H$\alpha$ luminosity, as

\begin{equation}\label{eq:Lbol}
    L_{\mathrm{bol, ~IMF}} \approx 138L_{H\alpha}.
\end{equation}

\begin{figure}[ht]
    \centering
    \includegraphics[width=0.48\textwidth,angle=270]{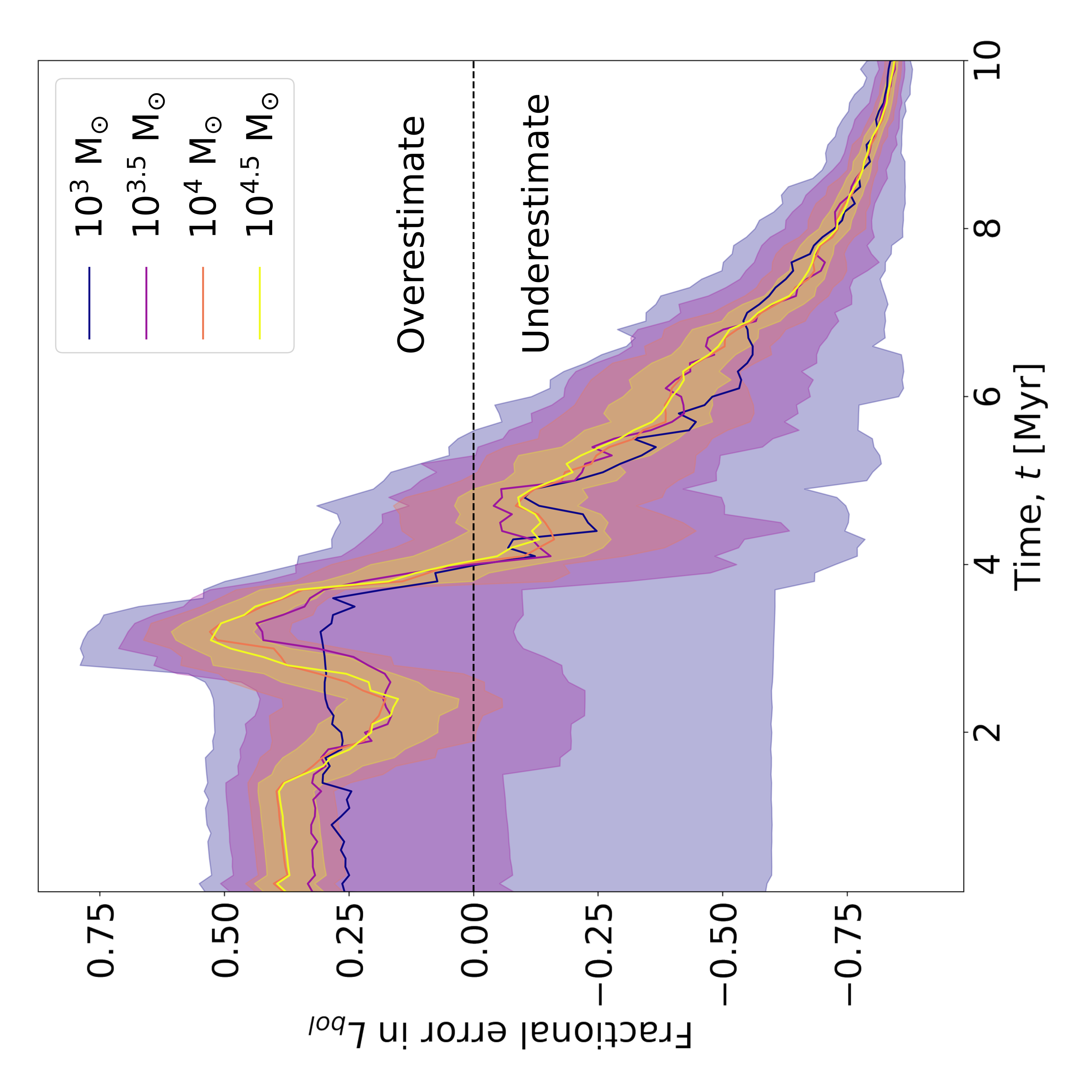}
    \caption{Fractional error when using Equation \ref{eq:Lbol} to determine $L_{\mathrm{bol}}$ compared to the true $L_{\mathrm{bol}}$ for synthesised stellar populations with different cluster masses as a function of time. To create these synthesised populations, we used \texttt{SLUG} with the median metallicity, extinction, and electron density found for the HII regions in this sample. The solid lines indicate the median of 100 simulations and the shaded region indicates the spread between the fifth and 95th percentiles for each cluster stellar mass. The dashed black line shows where the IMF-averaged estimate given by Equation \ref{eq:Lbol} is equivalent to the true bolometric luminosity from \texttt{SLUG}, $L_{\mathrm{bol, \texttt{SLUG}}}$. Above this line, Equation \ref{eq:Lbol} overestimates the true $L_{\mathrm{bol}}$ of these synthesised populations.}
    \label{fig:Lbol_simulated}
\end{figure}

We note that $L_{\mathrm{bol, ~IMF}}$ refers to the assumption that the stellar population fully samples the initial mass function (IMF). This relationship also assumes a continuous star formation history over a range of 0-100 Myr. However, individual star-forming regions are unlikely to fully sample the IMF and stellar age distribution. \citet{Lopez2014} and references therein find that, for low-mass regions, $L_{\mathrm{bol, ~IMF}}$ is likely to be an underestimate of the true bolometric luminosity, whilst for particularly young regions ($T\lesssim 5$ Myr), $L_{\mathrm{bol, ~IMF}}$ is likely to be an overestimate. In Figure \ref{fig:Lbol_simulated}, we use the stellar population synthesis code, \texttt{SLUG} (see Section \ref{sec:SPS}), to illustrate this using 100 synthesised stellar populations of varying stellar mass ($10^3$ - $10^{4.5} M_{\odot}$), where we calculate the fractional difference between using Equation \ref{eq:Lbol} to calculate the bolometric luminosity and the true bolometric luminosity of the synthesised stellar population. Positive values correspond to cases where Equation \ref{eq:Lbol} overestimates the true bolometric luminosity, and negative values to cases where it underestimates the true value. Here, the median fractional error for all masses is above 0 for populations younger than $\sim$ 4 Myr, indicating that using Equation \ref{eq:Lbol} is likely to overestimate $L_{\mathrm{bol}}$. However, for more evolved regions, Equation \ref{eq:Lbol} is more likely to be an underestimate, regardless of mass. We also see that there is a greater amount of stochastic variation for the low-mass, younger regions. In Section \ref{sec:cluster_prop} we compare the  $L_{\mathrm{bol}}$ values derived from Equation \ref{eq:Lbol} to those determined from a stochastically sampled stellar population, without assuming a fully sampled IMF, for the 30 HII regions analysed in this paper.

\subsection{Oxygen abundances}
\label{sec:oxygen}

\begin{figure}[h]

    \centering
    \includegraphics[width=0.4\textwidth,angle=270]{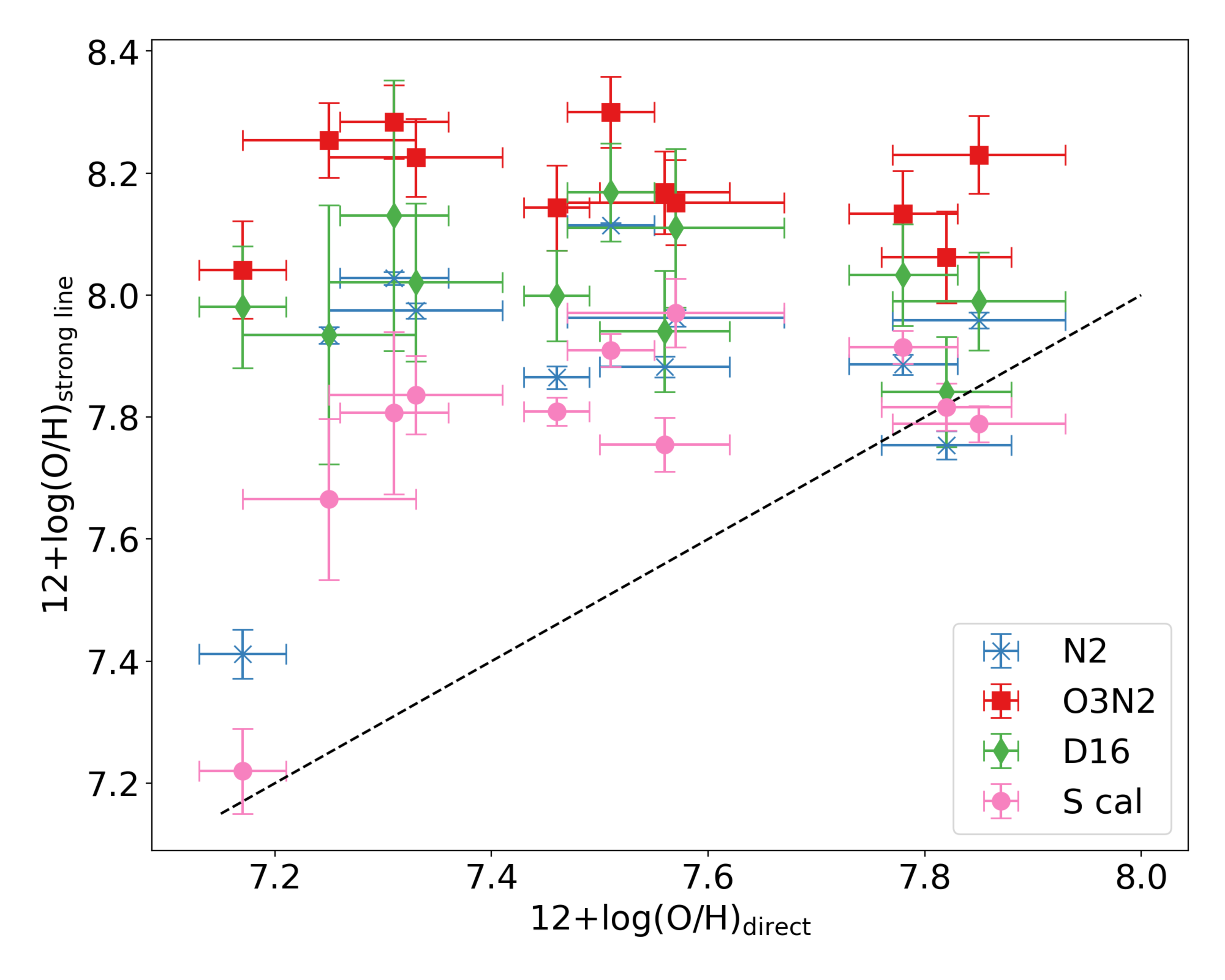}
    \caption{Plot of the oxygen abundances as determined from the strong-line calibration methods described in the Section \ref{sec:oxygen} against the direct method oxygen abundance for ten regions in J0921 (from \citetalias{J20}) and the Leo P HII region (taken from \citealt{Skillman_2013}), with the single Leo P region having the lowest $12+\log(\mathrm{O/H})_{\mathrm{direct}}$. The  dashed black line indicates where the derived abundances are equivalent to the direct, temperature-based values.}
    \label{fig:cdf o abundance}
    
\end{figure}

To derive oxygen abundances in HII regions as a measure of metallicity, the direct method is generally preferred (e.g. \citealt{Peimbert_2017,  Kewley2019b}). However, the MUSE wavelength range does not cover the most commonly used auroral line, [OIII]$\lambda$4363, and whilst it covers some of the others, these are typically very faint (\citealt{Curti_2016}). We do not achieve a high enough signal-to-noise (3$\sigma$ level) to measure these auroral lines for many of the HII regions. To make consistent metallicity estimates, we therefore use strong line calibrations to derive oxygen abundances.

The review of \citet{Kewley2019b} discusses various empirical and theoretical strong-line calibrations for deriving oxygen abundances, with often large discrepancies found between them. It is also well known that strong line abundances show a systematic offset to direct method abundances (e.g. \citealt{Pilyugin2016, Kumari2017}), and selecting a strong line calibration that minimises this offset is an oft-used strategy (e.g. \citealt{Yin2007} and references therein). For this study, we tested four different strong line calibrations, namely the S calibration method of \citet{Pilyugin2016}, the empirical N2 index and O3N2 index calibrations of \citet{peres_montero_2009}, and the theoretical calibration of \citet{Dopita}. These were selected both due to the wavelength coverage of the HII regions studied in this paper and for their being commonly used in recent studies relevant to this work (e.g. \citealt{Kreckel_2019, McLeod2021, McLeod2019, J20}). We compared the results of these different calibrations to available direct method values in order to determine which is the optimal method for our sample. For J0921, there are ten HII regions cross-matched to those in \citetalias{J20} with direct method abundances, and for Leo P we take the value given by \citet{McQuinn_2015}. 

In Figure \ref{fig:cdf o abundance}, we plot the oxygen abundances determined from the different calibrations against the direct method abundances for the ten corresponding regions in J0921 and the single region in Leo P. Here we see that the S calibration produces values of 12+log(O/H) (abbreviated to (O/H)$_{\mathrm{S cal}}$ in Table \ref{tab:results}), closest to those available that are determined via the direct method. The direct method and S calibration abundance values also show a similarly large spread for J0921 ($\sim$ 0.6 dex), indicating that this galaxy may be particularly chemically inhomogeneous, as discussed in \citetalias{J20}. The S calibration is an empirical calibration that relies on the following line ratios:

\noindent $N_2= \frac{F_{[NII]\lambda6548+\lambda6584}}{F_{H\beta}}$,\\ $S_2= \frac{F_{[SII]\lambda6716+\lambda6731}}{F_{H\beta}}$, \\ $R_3= \frac{F_{[OIII]\lambda4959+\lambda5007}}{F_{H\beta}}$,

\noindent where each of the flux terms are the Gaussian-fitted, continuum-subtracted and reddening-corrected integrated fluxes in each HII region for the corresponding emission line. This calibration is split into two branches;  a lower branch (for low-metallicity objects), where $\log(N_2) < -0.6$, given by Equation 7 in \citet{Pilyugin2016}; and an upper branch (for high-metallicity objects) where $\log(N_2)\geq-0.6$, given by Equation 6 in \citet{Pilyugin2016}. We find that the majority of our HII regions sit on the lower branch; that is to say, these are low metallicity regions.

We note that the use of different calibrations can, in some cases, produce conflicting findings, for example in searching for metallicity gradients and in quantifying values as a function of metallicity. We discuss some of these potential biases in Section \ref{sec:low_mass_z}, and in Appendices  \ref{appendix:j0921 comparison} and \ref{appendix:oxygen abundances}. Given the large discrepancies found between the various calibrations for our sample (up to $\sim$1 dex), and also for HII regions in the literature, deep spectroscopic observations of auroral lines will likely be necessary for future investigations of the dependence of stellar feedback on metallicity.

\section{Stellar population synthesis with SLUG}
\label{sec:SPS}

In the previous sections of this paper, we have mainly focused on the ionised gas within HII regions. However, a more complete understanding of stellar feedback requires some information on the feedback-driving stellar populations within these regions. When studying stellar populations, it is important to consider the limitations of the oft-used assumption that the stellar IMF is always fully sampled, so that population synthesis models can use IMF-averaged quantities. This is a poor assumption for low-mass galaxies, and for resolved, individual star-forming regions (\citealt{Silva2012} and references therein). For this reason, we have opted to use the stellar population synthesis (SPS) code \texttt{SLUG} (Stochastically Lighting Up Galaxies, \citealt{Silva2012, Krumholz_2015}, v2.0) to model stochastically sampled stellar populations within our sample of dwarf galaxies. In this section, we will discuss the use of this code for deriving the properties of the stellar populations within our HII regions.

For the purpose of this study, the main properties we are interested in deriving are the current stellar masses ($M_*$), ages ($T$), and bolometric luminosities ($L_{\mathrm{bol, ~\texttt{SLUG}}}$). In order to use \texttt{SLUG} to estimate these properties, we produce a library of $10^6$ simulations for each HII region in our sample, run in \texttt{SLUG}’s ‘cluster’ mode. For each library, we input the properties already derived for each HII region; namely the electron density, extinction, and metallicity. We note that each inputted metallicity is interpolated by \texttt{SLUG} to the nearest value available in the MESA Isochrones and Stellar Tracks (MIST) (\citealt{MIST}) and stellar atmosphere models, and therefore does not represent our full sample of metallicities. We use a \citet{Chabrier2005} IMF and a \citet{Calzetti2000} starburst attenuation curve with a factor of 2.27 between nebular and stellar attenuation (\citealt{Calzetti1994}). We sample cluster masses and ages from distributions:

\begin{equation}\label{eq:CMF}
    p_M(M_*) \propto M^{-1.5},  \qquad 100 <\frac{M_*}{M_\odot} \leq M_{\mathrm{max}} 
\end{equation}

\begin{equation}\label{eq:age distributioon}
    p_T(T) \propto T^{-0.5},  \qquad 10^6 <T [yr] \leq 10^7. 
\end{equation}

We initially set the maximum stellar mass, $M_{\mathrm{max}}$, to $5 \times 10^5 M_{\odot}$, similar to the maximum mass found in \citet{Barnes2021} and \citet{DellaBruna2022}, for example. Upon inspection of the resulting posterior distributions and considering its lower total stellar mass, we then limit this to $10^4 M_{\odot}$ for Leo P. The choice to limit the cluster age distribution between 1 to 10 Myr is based on the expectation that stellar populations older than 10 Myr are not typically associated with HII regions (e.g. \citealt{Kennicutt2012}).

The power laws of these distributions are selected to increase sampling of younger, less massive clusters, where there is a larger amount of stochastic variation in the resulting spectra and photometry. This sampling density is accounted for when deriving cluster properties, and therefore should not affect the physical results derived. 

We then follow the methods described in \cite{Krumholz2015_b}, whereby we use the \texttt{cluster\_slug}\footnote[1]{\scriptsize\url{https://slug2.readthedocs.io/en/latest/cluster_slug.html}} function to perform Bayesian inference of the cluster properties using a set of observed photometric values for each HII region and a physically motivated estimate of the prior probability given by:

\begin{equation}\label{eq:priors}
    p_{prior}(\textbf{x})\propto M_*^{-1}T^{-0.5}.
\end{equation}

Since high resolution imaging is not available for KKH046 and J0921, photometric values are calculated using a set of four custom filters to create mock photometric images using the MUSE data for each galaxy. For J0921 and Leo P, these images are created by collapsing the MUSE cubes over three broad wavelength ranges, 4650-5800 \AA, 5800-7000 \AA, and 7000-8200 \AA, and additionally 10 \AA ~around the redshifted H$\alpha$ emission line to create a mock narrow-band H$\alpha$ image. Whilst MUSE does cover redder wavelength ranges, we find that for these data cubes there may be residual emission from poor sky subtraction at wavelengths $>$ 8200 \AA. For the KKH046 observations with AO correction, we use the same four mock filters but with an inputted transmission of zero over the AO notch filter indicated in Figure \ref{fig:KKH046 spectrum}. Photometry values for these custom filters are also then outputted for each of our \texttt{SLUG} libraries, and compared to the observed values.

We take the median of the resulting posterior probability distribution functions (PDFs) as our estimated values for the mass, age and bolometric luminosity. To estimate the uncertainty on these values, we use the difference between the 84th and 16th percentiles (68\% confidence interval, CI).

\begin{table*}[t]
\caption[]{Properties of the 30 HII regions across the three dwarf galaxies in this sample.}
\normalsize
\centering
\begin{tabular}{llllllllll}
\toprule

    ID &  $r$ & $n_e$ & $A_V$ & $(O/H)_{S cal}$ & $M_*$ & $T$  & $P_{\mathrm{dir, ~IMF}}$ & $P_{\mathrm{dir,~ SLUG}}$ & $P_{\mathrm{ion}}$\\
       &   (pc)  & ($\mathrm{cm^{-3}}$) & (mag)  & & ($10^3 M_{\odot}$) & (Myr) & \multicolumn{3}{c}{log$_{10}$($P/k_B$) [K cm$^{-3}$]}         \\
    \midrule
    \multicolumn{10}{c}{\textbf{J0921}} \\
    \midrule
    1  &   70 ± 14 & 220 $\pm$ 100 & 0.55 $\pm$ 0.09 &  7.74 ± 0.2  & 0.7 $^{+1.0}_{-0.5}$ & 6.7 $^{+3}_{-2}$ & 2.7 $\pm$ 0.2 & 2.8 $\pm$ 0.2 & 6.6 $\pm$ 0.3\\
     2  &  213 ± 42 & 61 $\pm$ 9  & 0.44 $\pm$ 0.07 & 7.76 ± 0.04 & 4.8$^{+9.9}_{-3.4}$ & 6.2 $^{+2}_{-3}$ & 3.1 $\pm$ 0.2 & 2.9 $\pm$ 0.3 & 6.08 $\pm$ 0.06\\
     3  &  151 ± 30 & 130 $\pm$ 60  & 0.76 $\pm$ 0.1 & 7.71 ± 0.1 & 2.9$^{+5.5}_{-2.0}$ & 6.7 $\pm$ 2 & 2.8 $\pm$ 0.2 & 2.8 $\pm$ 0.3 & 6.4 $\pm$ 0.2\\
    4  &   98 ± 19 &  120 $\pm$ 50  &  0.71 $\pm$ 0.1 & 7.72 ± 0.1 & 1.4 $^{+2.1}_{-0.9}$ & 7.1 $\pm$ 3 & 2.6 $\pm$ 0.2 & 2.8 $\pm$ 0.3 & 6.4 $\pm$ 0.2\\
     5  &  148 ± 29 &  21 $\pm$ 5 & 0.65 $\pm$ 0.1 &  7.43 ± 0.08 & 2.1$^{+3.6}_{-1.4}$ & 7.2 $\pm$ 2 & 2.5 $\pm$ 0.2 & 2.6 $\pm$ 0.2 & 5.6 $\pm$ 0.1\\
     6  &  64 ± 12 & 74 $\pm$ 30  & 0.32 $\pm$ 0.05 &  7.55 ± 0.1 & 2.6 $^{+4.4}_{-1.8}$ & 7.1 $^{+2}_{-3}$ & 3.1 $\pm$ 0.2 & 3.6 $\pm$ 0.2 & 6.2 $\pm$ 0.1\\
     7  &  155 ± 31 & 43 $\pm$ 4 & 0.79 $\pm$ 0.1 & 7.79 ± 0.03 & 7 $^{+13}_{-5}$ & 6.8 $\pm$ 3 & 3.2 $\pm$ 0.2 & 3.4 $\pm$ 0.2 & 5.94 $\pm$ 0.04\\
     8  &  138 ± 33 & 46 $\pm$ 10  & 0.63 $\pm$ 0.1 & 7.72 ± 0.07 & 3.3 $^{+5.6}_{-2.3}$ & 7.0 $\pm$ 3 & 2.9 $\pm$ 0.2 & 3.0 $\pm$ 0.3 & 6.0 $\pm$ 0.1\\
    9  &  169 ± 30 & 45 $\pm$ 5  & 0.63 $\pm$ 0.1&  7.82 ± 0.04 & 6 $^{+13}_{-4}$ & 6.5 $\pm$ 3 & 3.2 $\pm$ 0.2 & 3.1 $\pm$ 0.3 & 5.95 $\pm$ 0.05\\
    10  &  95 ± 19 & 77 $\pm$ 9  & 0.85 $\pm$ 0.1 & 7.91 ± 0.03 & 2.4 $^{+4.1}_{-1.7}$ & 6.4 $\pm$ 3 & 3.4 $\pm$ 0.2 & 3.1 $\pm$ 0.3 & 6.19 $\pm$ 0.05\\
    11 &  94 ± 18 &  200 $\pm$ 70  & 0.59 $\pm$ 0.1 & 7.96 ± 0.1 & 1.4 $^{+2.1}_{-0.9}$ & 6.7 $^{+3}_{-2}$ & 2.8 $\pm$ 0.2 & 2.8 $\pm$ 0.3 & 6.6 $\pm$ 0.2\\
    12 &  158 ± 31 & 38 $\pm$ 4 & 0.67 $\pm$ 0.1 & 7.81 ± 0.02 & 7 $^{+15}_{-5}$ & 6.4 $\pm$ 3 & 4.0 $\pm$ 0.2 & 3.3 $\pm$ 0.3 & 5.89 $\pm$ 0.04\\
    13 &  90 ± 18 &  78 $\pm$ 40  &  0.96 $\pm$ 0.2 & 7.81 ± 0.1 & 2.0 $^{+3.1}_{-1.3}$ & 7.3 $\pm$ 3 & 2.2 $\pm$ 0.2 & 2.9 $\pm$ 0.2 & 6.2 $\pm$ 0.2\\
    14 &  68 ± 13 & 58 $\pm$ 40 &  0.61 $\pm$ 0.1 & 7.85 ± 0.1 & 1.1 $^{+1.6}_{-0.7}$ & 7.2 $\pm$ 3 & 2.4 $\pm$ 0.2 & 3.0 $\pm$ 0.3 & 6.1 $\pm$ 0.3\\
    15 &  229 ± 45 & 91 $\pm$ 10 &  0.60 $\pm$ 0.1& 7.91 ± 0.03 & 14 $^{+28}_{-9}$ & 6.7 $\pm$ 2 & 4.2 $\pm$ 0.2 & 3.2 $\pm$ 0.3 & 6.26 $\pm$ 0.05\\
    16 & 153 ± 30 & 77 $\pm$ 20  &  1.1 $\pm$ 0.2&  7.97 ± 0.06 & 7 $^{+14}_{-5}$ & 7.0 $\pm$ 2 & 2.6 $\pm$ 0.2 & 3.1 $\pm$ 0.2 & 6.2 $\pm$ 0.1\\
     17 & 88 ± 17 & 29 $\pm$ 10 & 0.80 $\pm$ 0.1 & 7.67 ± 0.1 & 1.9 $^{+2.9}_{-1.2}$ & 7.2 $\pm$ 2 & 2.3 $\pm$ 0.2 & 3.0 $\pm$ 0.3 & 5.8 $\pm$ 0.2\\
     18 &  178 ± 35 & 100 $\pm$ 20 &0.67 $\pm$ 0.1& 7.84 ± 0.06 & 5.1 $^{+9.6}_{-3.5}$ & 6.8 $\pm$ 4 & 3.5 $\pm$ 0.2 & 3.0 $\pm$ 0.2 & 6.3 $\pm$ 0.1\\
     19 & 43 ± 8.0 & 12 $\pm$ 4 &1.0 $\pm$ 0.2&  8.02 ± 0.07 & 3.3 $^{+4.6}_{-2.1}$ & 7.3 $^{+2}_{-3}$ & 2.2  $\pm$ 0.2 & 3.8 $\pm$ 0.2 & 5.4 $\pm$ 0.2\\
     20 & 97 ± 19 & 260 $\pm$ 100 &0.99 $\pm$ 0.2 & 7.79 ± 0.1 & 1.7$^{+2.6}_{-1.1}$ & 7.2 $\pm$ 3 & 2.6 $\pm$ 0.2 & 2.8 $\pm$ 0.2 & 6.7 $\pm$ 0.2\\
    21 &  98 ± 19 & 230 $\pm$  60 &0.74 $\pm$ 0.1 & 7.96 ± 0.08 & 2.0$^{+3.1}_{-1.3}$ & 7.5 $\pm$ 2 & 2.7 $\pm$ 0.2 & 2.9 $\pm$ 0.2 & 6.7 $\pm$ 0.1\\
    22 &  105 ± 21 & 46 $\pm$ 10  & 0.79 $\pm$ 0.1 & 7.58 ± 0.09 & 2.6 $^{+4.4}_{-1.7}$ & 7.3 $\pm$ 2 & 2.8 $\pm$ 0.2 & 2.9 $\pm$ 0.2 & 6.0 $\pm$ 0.1\\
    23 &  210 ± 42 & 58 $\pm$ 7 &0.55 $\pm$ 0.09 & 7.62 ± 0.05 & 7 $^{+14}_{-5}$ & 7.1 $\pm$ 3 & 3.4 $\pm$ 0.2 & 3.0 $\pm$ 0.3 & 6.06 $\pm$ 0.05\\
    \midrule
    \multicolumn{9}{c}{\textbf{KKH046}} \\
    \midrule
    1 &  184 ± 36 &  36 $\pm$ 7 & 0.79 $\pm$ 0.1 & 7.69 ± 0.09 & 5 $^{+13}_{-4}$  & 5.5 $\pm$ 3 & 3.2 $\pm$ 0.2 & 2.9 $\pm$ 0.3 & 5.85 $\pm$ 0.09\\
     2 &  249 ± 50 &  5.3 $\pm$ 0.9 & 0.71 $\pm$ 0.1 & 7.92 ± 0.05 & 7 $^{+14}_{-5}$ & 5.4 $\pm$ 3 & 3.0 $\pm$ 0.2 & 2.8 $\pm$ 0.2 & 5.02 $\pm$ 0.07\\
    3 &  208 ± 42 &  17 $\pm$ 2 & 0.61 $\pm$ 0.1  & 7.96 ± 0.04 & 4.2 $^{+7.2}_{-2.8}$ & 4.6 $\pm$ 3 & 3.2 $\pm$ 0.2 & 2.8 $\pm$ 0.3 & 5.52 $\pm$ 0.06\\
     4 &  150 ± 30 & 29 $\pm$ 7 & 0.70 $\pm$ 0.1 &  7.92 ± 0.06 & 3.7$^{+7.7}_{-2.5}$ & 4.8 $\pm$ 3 & 3.4 $\pm$ 0.2 & 3.0 $\pm$ 0.3 & 5.8 $\pm$ 0.1\\
     5 &   87 ± 17 &  51 $\pm$ 22& 0.63 $\pm$ 0.1  & 7.94 ± 0.2 & 2.6 $^{+4.4}_{-1.8}$ & 6.1 $\pm$ 3 & 3.1 $\pm$ 0.2 & 3.3 $\pm$ 0.2 & 6.0 $\pm$ 0.2\\
     6 &  274 ± 55 &  72 $\pm$ 5& 0.56 $\pm$ 0.09  & 7.96 ± 0.03 & 9$^{+16}_{-5}$ & 4.5 $^{+2}_{-3}$ & 3.5 $\pm$ 0.2 & 2.9 $\pm$ 0.3 & 6.16 $\pm$ 0.03\\
     \midrule
    \multicolumn{9}{c}{\textbf{Leo P}} \\
    \midrule
    1  &  22 ± 4.0 &  91 $\pm$ 30 & 0.28 $\pm$ 0.01  & 7.24 ± 0.07 & 0.5 $^{+0.8}_{-0.4}$ & 5.4 $^{+2}_{-1}$ & 3.9 $\pm$ 0.2 & 4.3 $\pm$ 0.1 & 6.3 $\pm$ 0.1\\
\bottomrule
\end{tabular}
\normalfont
\label{tab:results}
\tablefoot{
     Columns correspond to (1) region ID, (2) radius of the region, (3) electron density, (4) visual extinction, (5) oxygen abundance, (6) cluster stellar mass, (7) cluster age, (8) direct radiation pressure assuming a fully sampled IMF, (9) direct radiation pressure using a stochastically sampled stellar population, and (10) the pressure of the ionised gas.
    }
\end{table*}

\subsection{Star cluster properties}
\label{sec:cluster_prop}

\begin{figure*}[h]
    \centering
    \includegraphics[width=\textwidth]{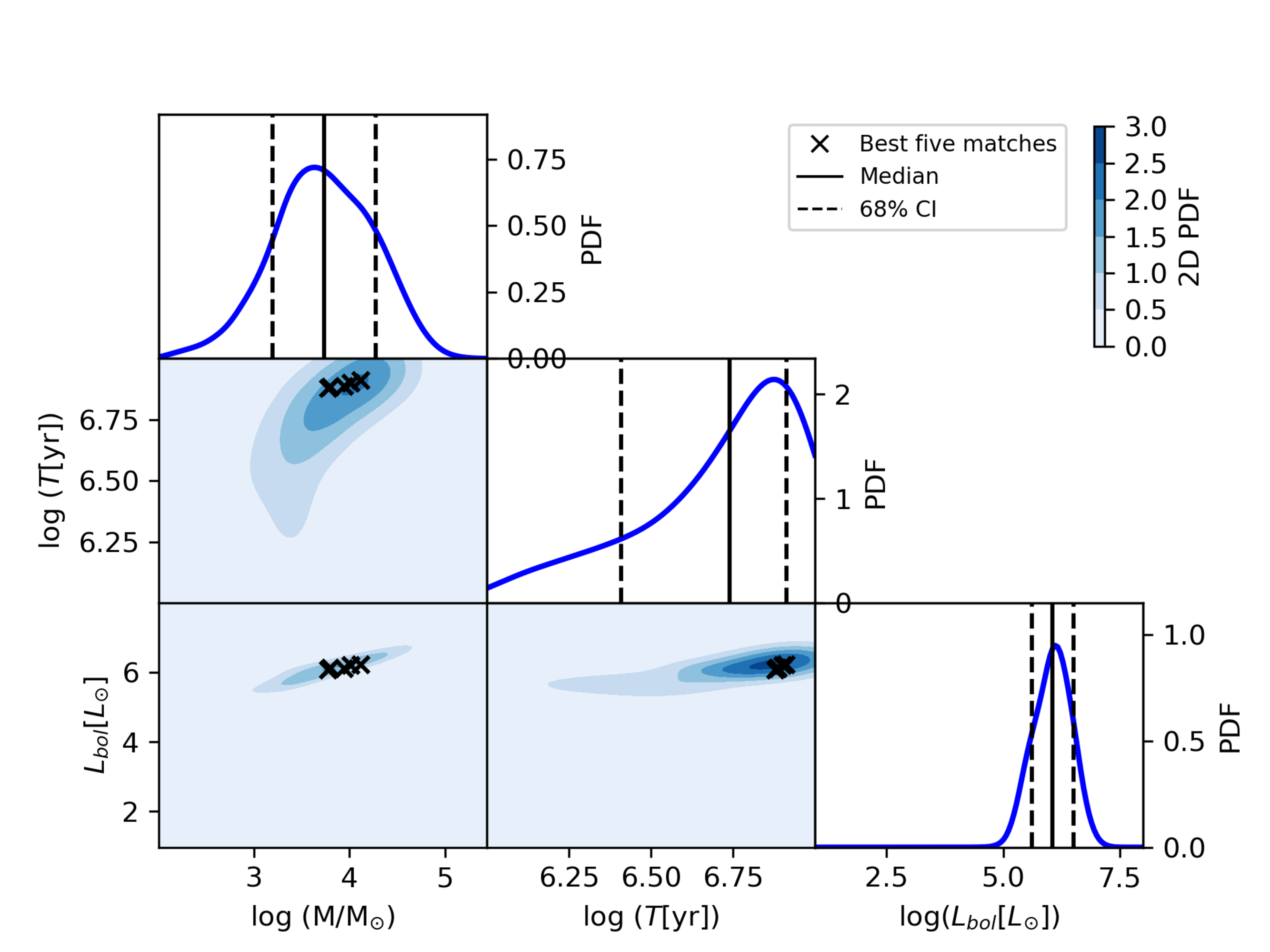}
    \caption {Corner plot for KKH046 region 1 showing the 1D posterior probability distribution functions (PDFs) for the stellar mass ($\log M_*$),  age ($\log T$) and bolometric luminosity ($\log L_{\mathrm{bol}}$) of this region in the top, centre, and bottom right panels, respectively. The vertical solid and dashed lines indicate the median values and their 68\% confidence interval (CI), respectively. The bottom left panels show the joint posterior PDFs, with the intensity indicating the probability density. The five crosses indicate the five best photometric matches in the library to the observed photometry. All PDFs are normalised to have unit integral.}
    \label{fig:triangle plot}
\end{figure*}

In Figure \ref{fig:triangle plot}, we show the posterior PDFs for the mass, age and bolometric luminosity of KKH046 region 1 as an example of the capabilities of \texttt{cluster\_slug}. From this plot, and from similar plots of all other regions, we see that in the majority of cases, the posterior PDFs returned by \texttt{cluster\_slug} have a well-defined peak within the range of values tested for the cluster mass and bolometric luminosity. In the case where the distribution is bimodal, the median typically sits somewhere between these peaks, as in \citet{Krumholz_2015}. However, the ages of these clusters are not as well constrained with these observations, likely due to the limited wavelength range of the mock photometry.

We find that the star clusters powering these HII regions have estimated stellar masses ($M_*$) within the range of $500$ to $10^4 M_{\odot}$, and estimated ages between $4.5$ to $7.3$ Myr. In Figure \ref{fig:triangle plot} we also plot the joint posterior PDF for the cluster properties, along with the five best photometric matches in this library to the observed photometry of KKH046 region 1. We find that for all regions, a minimum of two best matches lies within the 3$\sigma$ contour level of the joint PDFs. 

\begin{figure}[h]
    \centering
    \includegraphics[width=0.35\textwidth,angle=270]{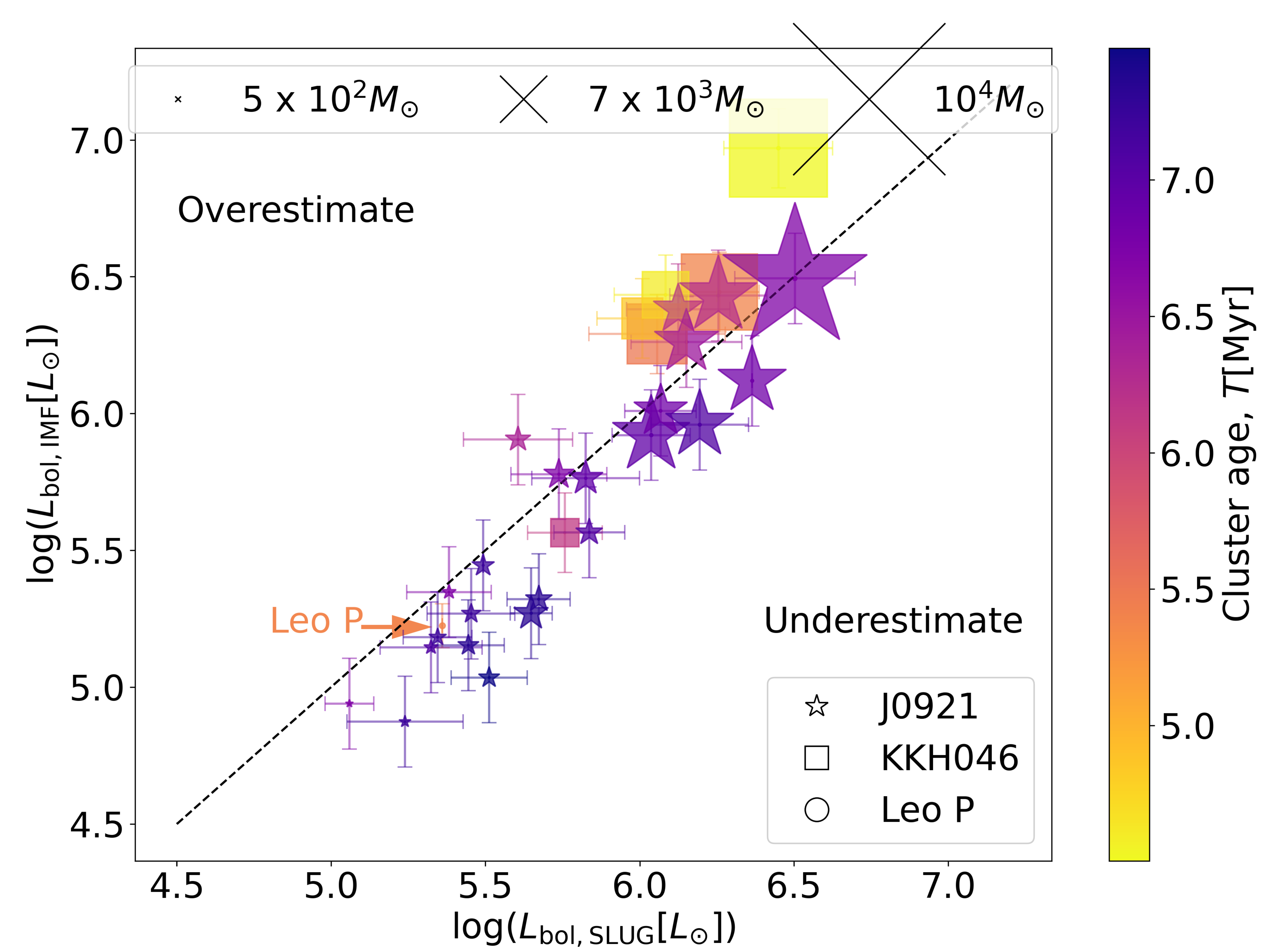}
    \caption {Comparison plot for the two methods used to calculate the bolometric  luminosity for the 30 HII regions in this sample, where for $L_{\mathrm{bol, ~IMF}}$ we have assumed a fully sampled IMF, and for $L_{\mathrm{bol, \texttt{SLUG}}}$ we have used the median of the posterior PDF returned by \texttt{SLUG}. The colours indicate the age of the regions, and the size of each point is a representation of their mass. The dashed black line indicates where these values are equivalent; points above this line show where $L_{\mathrm{bol, IMF}}$ is greater than $L_{\mathrm{bol, \texttt{SLUG}}}$ - that is to say, where assuming a fully sampled IMF with Equation \ref{eq:Lbol} overestimates the bolometric luminosity. The Leo P data point has been labelled for visual clarity.} 
    \label{fig:p dir comparison}
\end{figure}

In this paper, we have determined two different values for the bolometric luminosity of each HII region, the first using the approximation given by Equation \ref{eq:Lbol} ($L_{\mathrm{bol, ~IMF}}$) and the second estimated with \texttt{SLUG} ($L_{\mathrm{bol, ~\texttt{SLUG}}}$). We plot in Figure \ref{fig:p dir comparison} a comparison of these two methods. These data points are coloured by age, and their size corresponds to their mass. We find that for the majority of our regions, $L_{\mathrm{bol, IMF}}< L_{\mathrm{bol, \texttt{SLUG}}}$. This follows from our finding that these regions are low mass ($M_* \lesssim 6 \times 10^4 M_{\odot}$) and not particularly young ($T \gtrsim 4.5$ Myr). The uncertainty in using Equation \ref{eq:Lbol} to estimate the bolometric luminosity, as discussed in Section \ref{sec:lum}, is therefore likely to be dominated by the assumption that the IMF is fully sampled, leading to an underestimation of the bolometric luminosity, and consequently the direct radiation pressure.

An additional useful parameter returned by \texttt{SLUG} is the mass of the most massive star currently present within the stellar population. This parameter is of particular interest for our investigation of Leo P, since pre-existing studies (\citealt{Evans2019}; \citealt{Telford2021, Telford2023}) have identified a single O-type star within this HII region of mass $22^{+7}_{-5} M_{\odot}$ and age $7^{+4}_{-3}$ Myr (\citealt{ Telford2021}). To make an estimate from \texttt{SLUG}, we create 20 libraries each of 1000 simulations and find the best photometric match for each library to the star cluster in the Leo P HII region. We then record the mass and age of the most massive star in that best match. We find a median and 68\% confidence interval of $23^{+4}_{-7} M_{\odot}$ for the mass of the most massive star and an age of $8^{+2}_{-1}$ Myr. This shows that our photometric method yields results consistent with spectroscopic studies.

\section{Quantifying the mechanisms of stellar feedback}
\label{sec:quantifying}

\begin{figure*}[t]
    \centering
    \begin{subfigure}[b]{0.48\textwidth}
        \centering
        \includegraphics[width=\textwidth,angle=270]{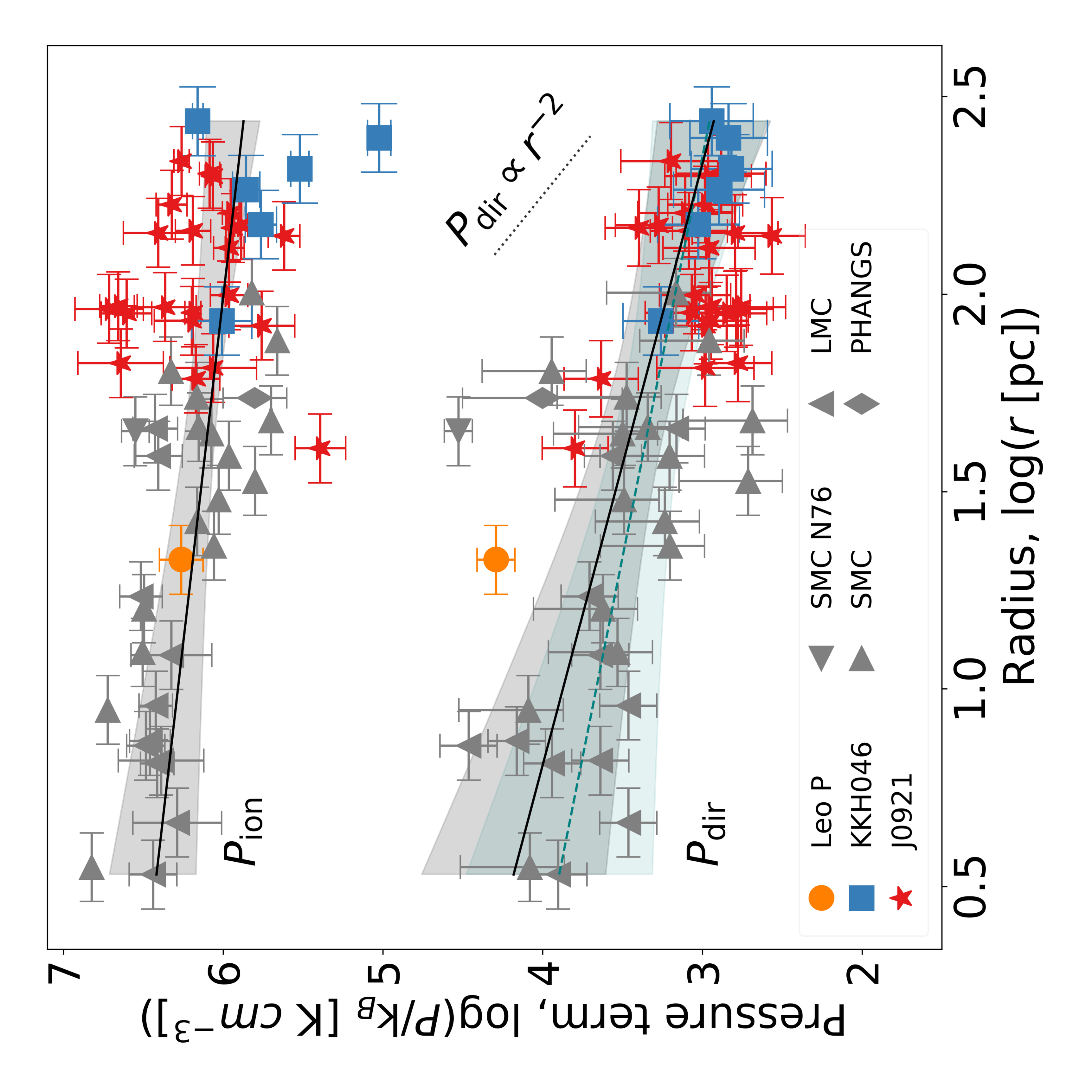}
        \caption{}
        \label{fig:p vs r}
    \end{subfigure}
    \begin{subfigure}[b]{0.51\textwidth}
        \centering
        \includegraphics[width=\textwidth,angle=270]{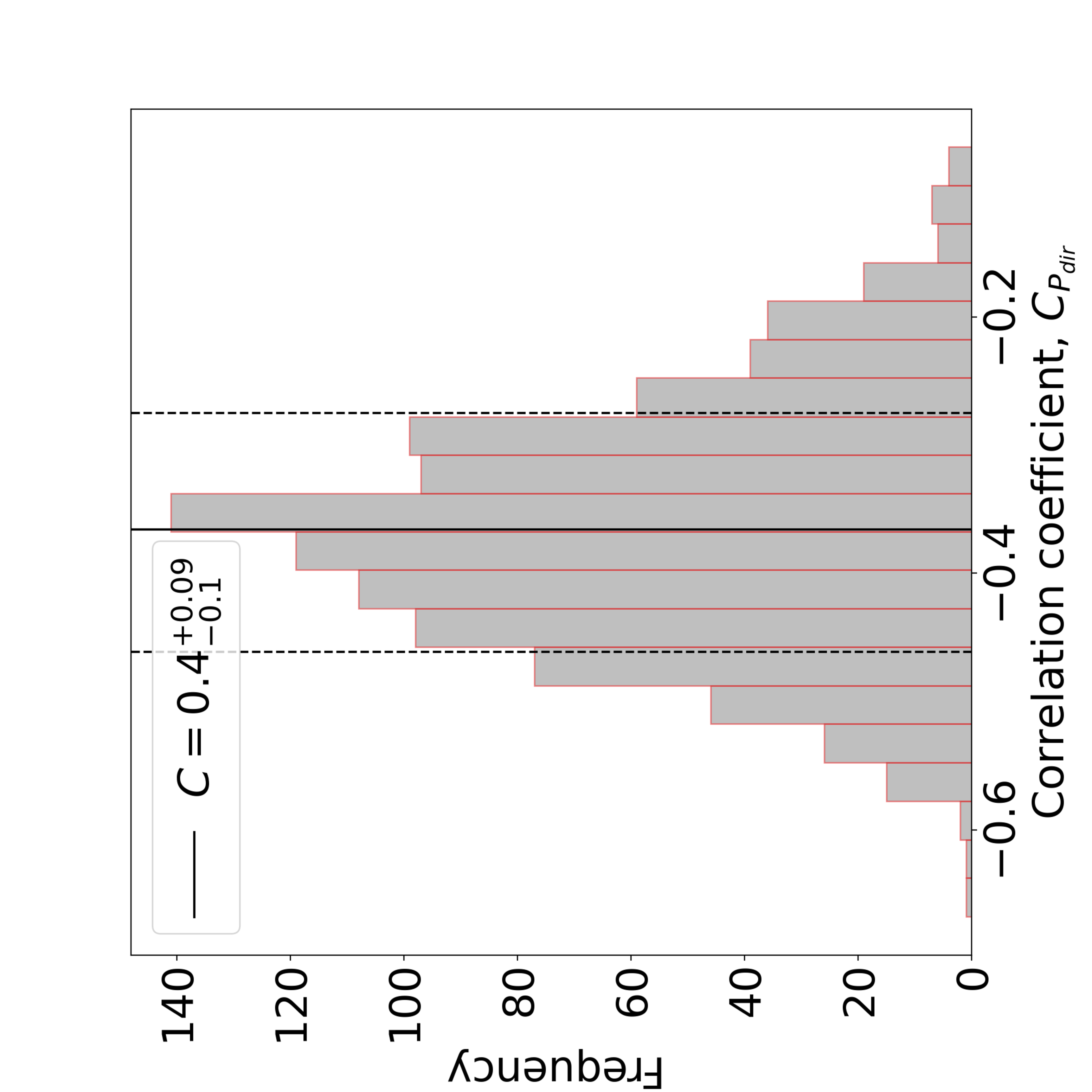}
        \caption{}
        \label{fig:bootstraps}
    \end{subfigure}
    \caption {Panel (a): Photoionisation pressure ($P_{\mathrm{ion}}$) and direct radiation pressure ($P_{\mathrm{dir}}$) of the HII regions in various galaxies (see legend) plotted as a function of region size. The $P_{\mathrm{ion}}$ values occupy the top half of the plot, and $P_{\mathrm{dir}}$ values the bottom half. SMC points are taken from \citet{Lopez2014},  SMC N76 from McLeod et al. (in prep), and the LMC points from \citet{McLeod2019}. We note that uncertainties in the radii values are not given in the literature, and we therefore assume a generous 20\%. We also plotted the median values from the PHANGS-MUSE survey as given in \citet{Barnes2021} by the grey diamond marker. The solid black lines indicate the linear best fits for $P_{\mathrm{ion}}$ and $P_{\mathrm{dir, ~\texttt{SLUG}}}$, with shaded grey areas for the 95\% confidence regions. The dashed teal line and shaded area depict the linear fit when using $P_{\mathrm{dir, ~IMF}}$. The black dotted line indicates the $r^{-2}$ proportionality from Equation \ref{eq:Pdir} used to calculate $P_{\mathrm{dir}}$. Panel (b): Correlation coefficient analysis for the $P_{\mathrm{dir, ~\texttt{SLUG}}} - \log_{10}(r)$ relation (see Section \ref{sec:Discussion}). The histogram shows the probability density distribution of the Pearson correlation coefficient using the bootstrapping and perturbation method from \citet{Curran2014}. The solid black line is the median correlation coefficient, $C$, found for this relation and the dashed lines show the 16th and 84th percentiles used to determine the uncertainty. }
    \label{fig:evolution}
\end{figure*}

\begin{figure}
    \centering
    \includegraphics[width=0.5\textwidth,angle=270]{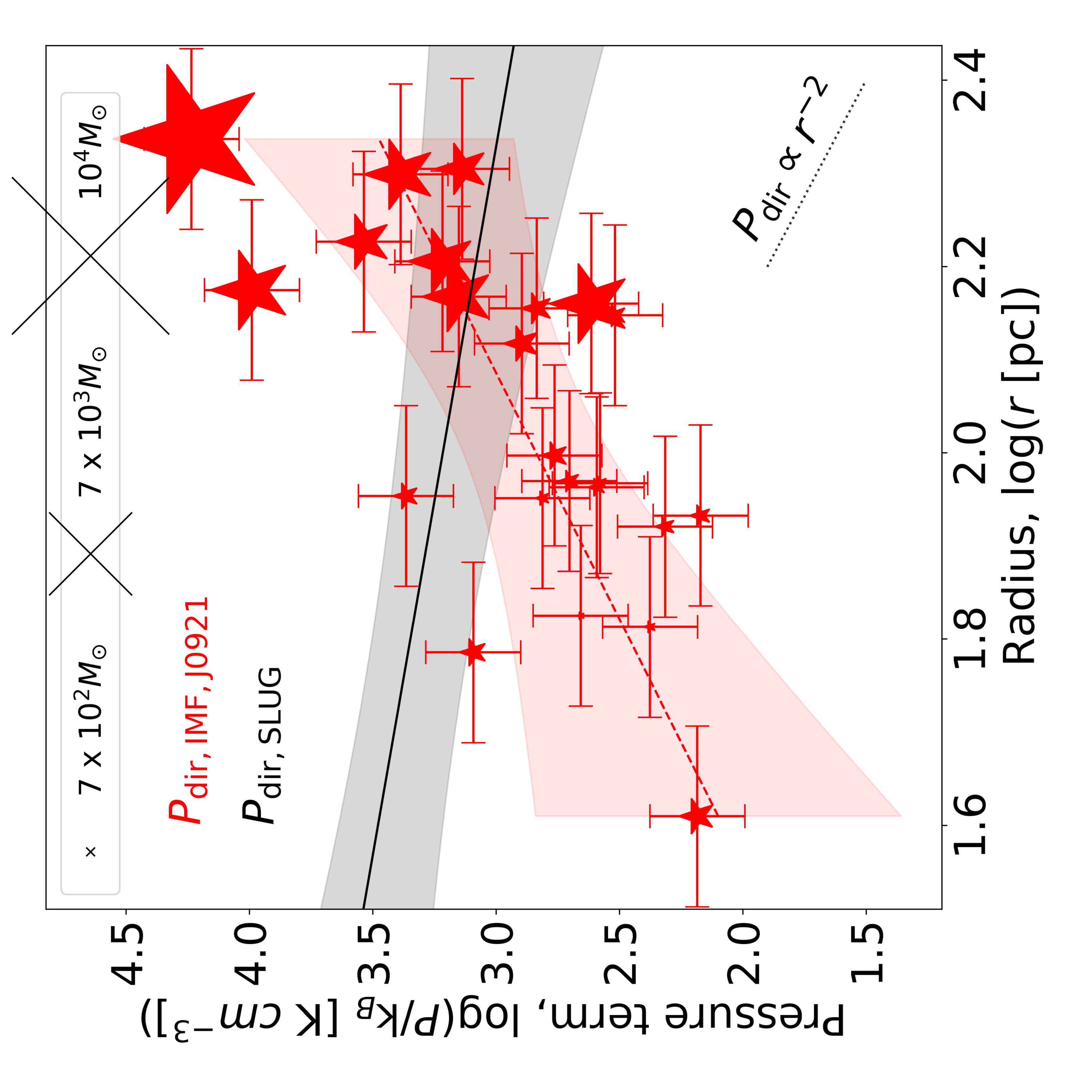}
    \caption{Here we plot only the $P_{\mathrm{dir, ~IMF}}$ values for J0921 against their HII region radii with red star-shaped markers sized according to their stellar mass. The solid black line and shaded  area indicate the $P_{\mathrm{dir, ~\texttt{SLUG}}}$ fit from Figure \ref{fig:p vs r}. The dashed red line and shaded area indicate the $P_{\mathrm{dir, ~IMF}}$ fit for the HII regions in J0921 only.}
    \label{fig:J0921_P_vs_r}
   
\end{figure}

\begin{figure*}[h]
    \centering
    \begin{subfigure}[b]{0.48\textwidth}
        \centering
        \includegraphics[width=\textwidth,angle=270]{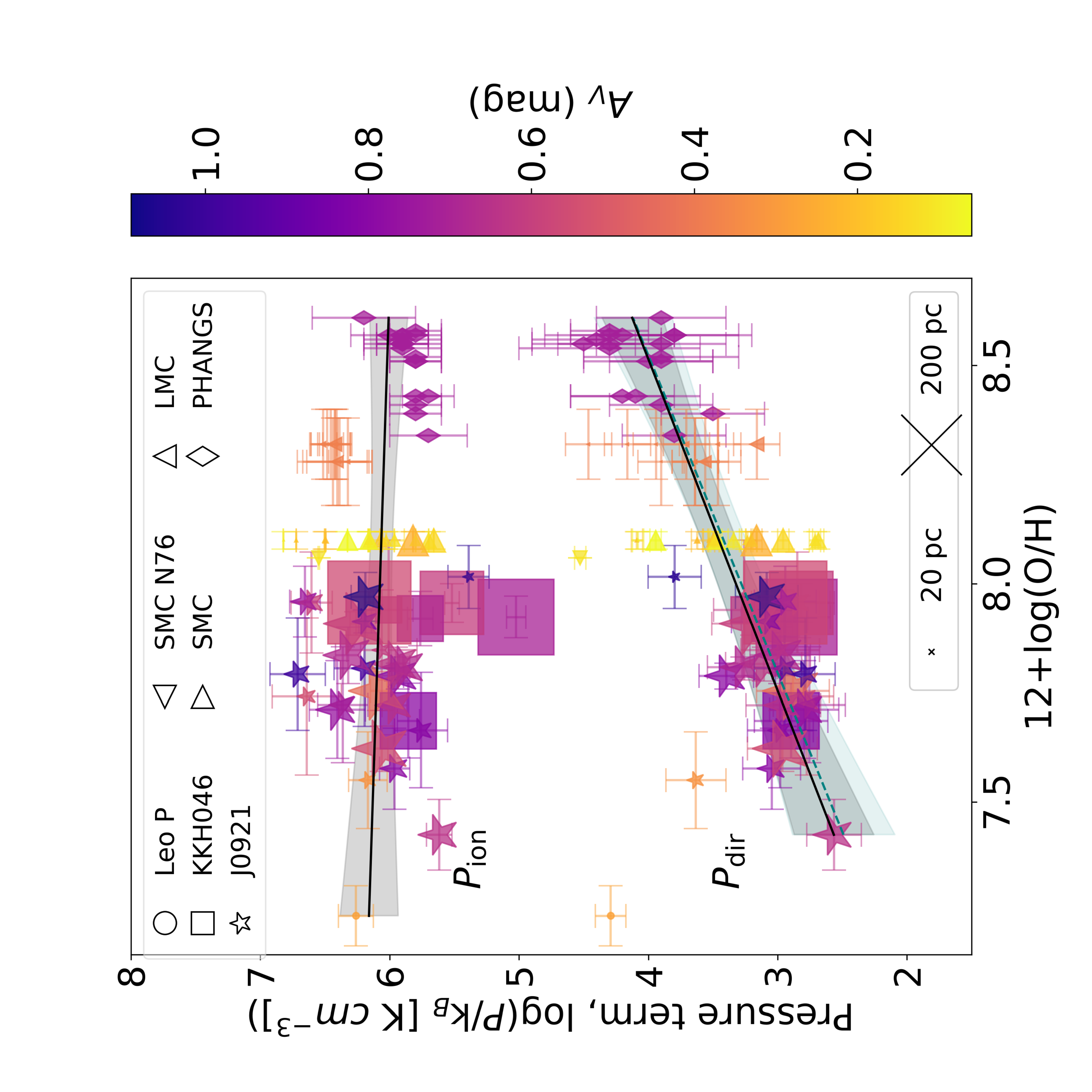}
        \caption{}
        \label{fig:p vs z}
    \end{subfigure}
    \begin{subfigure}[b]{0.48\textwidth}
        \centering
        \includegraphics[width=\textwidth,angle=270]{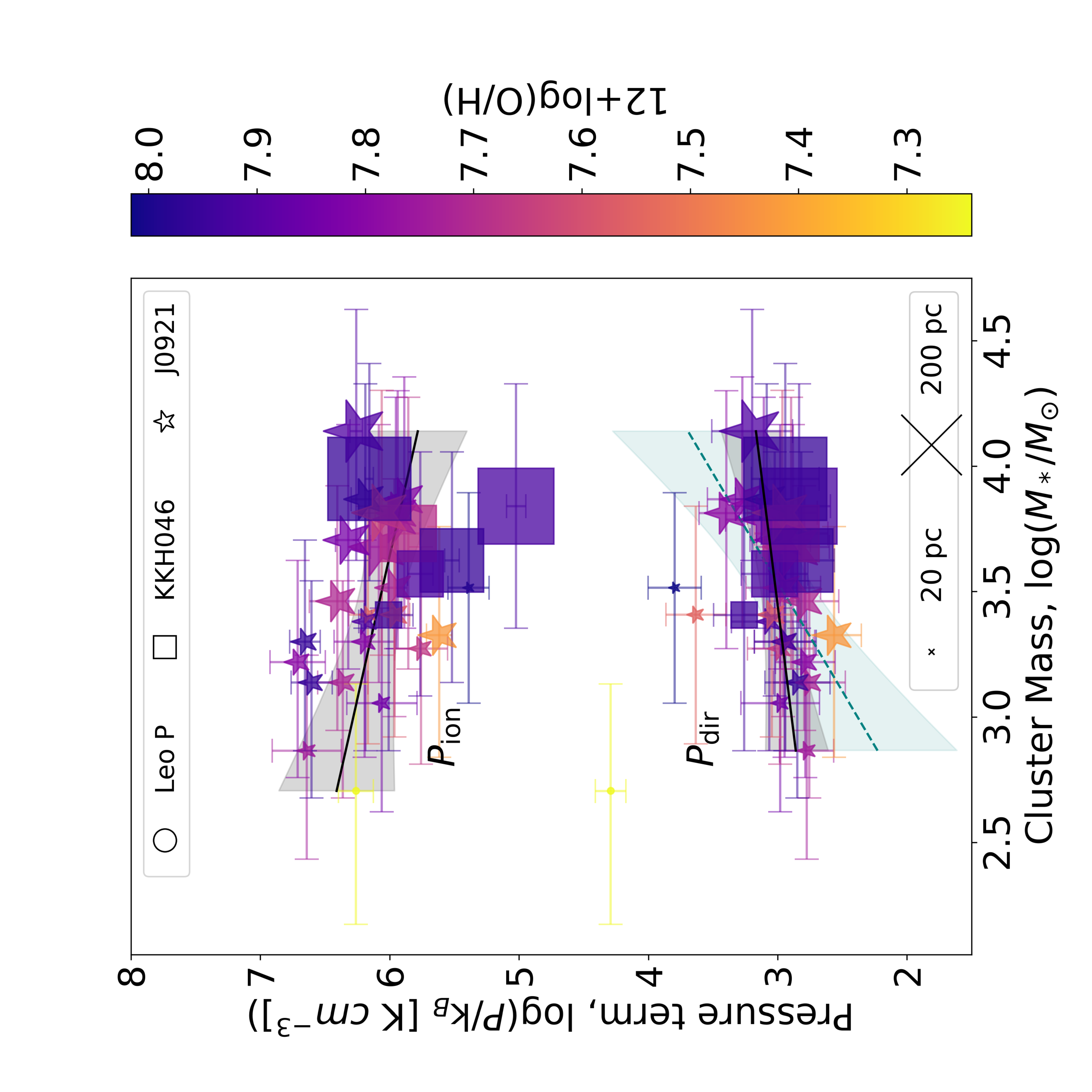}
        \caption{}
        \label{fig:p vs m}
    \end{subfigure}
    \caption{ Panel (a): Similar to Figure \ref{fig:p vs r}, but with the pressure components plotted as a function of their oxygen abundance, 12+log(O/H), with all markers coloured by their extinction and sized according to the HII region radii, where we take the median radii and extinction for the PHANGS data points (\citealt{Barnes2021}). Panel (b): Similar to panel (a), but with only the 30 HII regions from the three dwarfs in this sample plotted against their cluster mass and coloured by their oxygen abundance.}
    \label{fig:low mass and Z}
\end{figure*}

In this work, we focus on two pre-SN stellar feedback mechanisms, namely photoionisation and direct radiation pressure. These mechanisms can be quantified by their corresponding pressure terms, that is, the  thermal pressure of the ionised gas and the mechanical direct radiation pressure resulting from the momentum imparted by the stellar light output (as in e.g. \citealt{Lopez2014, McLeod2021}). There exist other pre-SN mechanisms of interest, including stellar winds and the dust-processed radiation pressure. We do not attempt to derive stellar wind pressure, since this derivation often depends on models that may not be valid at the low-metallicity regime (e.g. \citealt{Rickard2022} and references therein). Follow-up X-ray observations of our sample would allow us to better assess the hot gas pressure attributed to shock-heating by stellar winds (and SNe), and follow-up infrared observations would enable us to quantify the dust-processed radiation pressure (e.g. as in the PHANGS-JWST programme for massive star-forming galaxies, \citealt{Lee2023}). 

Despite a number of feedback-related pressure terms being outside the scope of this work, it is likely that the ionised gas pressure (together with stellar winds) will be more dominant within this sample, as previously found for evolved HII regions (e.g. \citealt{McLeod2021, Lopez2014}).

\subsection{Direct radiation pressure}
\label{sec:Pdir}

We determine the total volume-averaged direct radiation pressure, $P_{\mathrm{dir}}$, as

\begin{equation}\label{eq:Pdir}
    P_{\mathrm{dir}}=\frac{3L_{\mathrm{bol}}}{4\pi r^2 c},
\end{equation}

\noindent where $c$ is speed of light and $r$ is the radius of the HII region.

Two different values for the direct radiation pressure of each HII region are calculated, the first using the approximation for $L_{\mathrm{bol,~IMF}}$ given by Equation \ref{eq:Lbol} and the second using the value of $L_{\mathrm{bol,~  \texttt{SLUG}}}$ estimated with \texttt{SLUG}, referred to as $P_{\mathrm{dir, ~IMF}}$ and $P_{\mathrm{dir, ~\texttt{SLUG}}}$, respectively. The overall effect of using $L_{\mathrm{bol, ~\texttt{SLUG}}}$ instead of $L_{\mathrm{bol, ~IMF}}$ is that it shifts the direct radiation pressure values up for most of the HII regions in this sample.

For visual clarity, we show only the $P_{\mathrm{dir, ~\texttt{SLUG}}}$ points in all of our subsequent figures. However, we show the corresponding best-fit lines found using $P_{\mathrm{dir, ~IMF}}$ in some figures by a dashed teal line for comparison.

\subsection{Pressure of the ionised gas}
\label{sec:Pion}

We compute the ionised gas pressure from the ideal gas law, 

\begin{equation}\label{eq:Pion}
    P_{\mathrm{ion}}=(n_e + n_{\mathrm{H}} + n_{\mathrm{He}}) k_B T_e \approx 2  n_e k_B T_e,
\end{equation}

\noindent where $n_e$, $n_{\mathrm{H}}$ and $n_{\mathrm{He}}$ are the electron, hydrogen, and helium number densities, respectively, and $k_B$ is the Boltzmann constant. As with our derivation of the electron density, $n_e$, in Equation \ref{eq:density}, we assume that the temperature of the HII regions is 10$^4$ K. We also assume that these regions are neutral and that helium is singly ionised, so that $n_e + n_H + n_{He} \approx 2 n_e$.

\section{Discussion}
\label{sec:Discussion}

We now explore our findings and their implications for the relationships between stellar feedback and environmental properties in HII regions. We compare to studies with similar spatial resolution, including from the LMC (\citealt{McLeod2019}; HII region radii of 3.4-45.7 pc), the SMC (\citealt{Lopez2014}; HII region radii ranging from 10-112 pc), and the PHANGS-MUSE survey (\citealt{Barnes2021}; spatial resolution ranging from 50-100 pc) in order to investigate a wider range of physical properties in star-forming galaxies. The range of $\approx 7 - 8$ in $12 + \log(\mathrm{O/H})$ in our sample allows us to expand these studies into the less studied, low-metallicity regime.

The $P_{\mathrm{ion}}$ values for the regions in the LMC from \citet{McLeod2019} are determined using the same methods as described in Section \ref{sec:Pion}. However, the $P_{\mathrm{dir}}$ values are determined using the ionising photon flux, $Q_{0,*}$, of the stellar populations identified within the regions. \citet{McLeod2019} state that if Equation \ref{eq:Pdir} and $L_{\mathrm{bol,~ IMF}}$ from Equation \ref{eq:Lbol} is used, the $P_{\mathrm{dir}}$ values would be around an order of one magnitude higher. However, since these regions are young, it is likely that $L_{\mathrm{bol,~ IMF}}$ would be an overestimate. If the $P_{\mathrm{dir}}$ values for the LMC regions were higher, this would only strengthen the trends observed in the following discussions. For the HII regions in the SMC taken from \citet{Lopez2014}, and for the region SMC N76 (McLeod et al. in prep), the pressure terms are determined using the same methods as described in Section \ref{sec:quantifying}, but with a fully sampled IMF assumed when determining $L_{\mathrm{bol}}$, increasing their uncertainty. For the PHANGS-MUSE sample, we take the median pressure terms for each of the 19 galaxies in \citet{Barnes2021}. In \citet{Barnes2021}, two values are calculated for both $P_{\mathrm{dir}}$ (which is called $P_{\mathrm{rad}}$ in \citealt{Barnes2021}) and for $P_{\mathrm{ion}}$ (which is called $P_{\mathrm{therm}}$ instead). We take the minimum $P_{\mathrm{dir}}$ values ($P_{\mathrm{rad,~ min}}$) and the maximum values of $P_{\mathrm{ion}}$ ($P_{\mathrm{therm,~ max}}$), since the methods used to determine these are consistent with those used for the 30 HII regions in this study, and with the other literature studies described above.  We note that taking the minimum values for both pressure terms, or the maximum values for both pressure terms, would strengthen the findings discussed in Section \ref{sec:low_mass_z}. A lower conversion of 88 between $L_{H\alpha}$ and $L_{\mathrm{bol}}$ is used instead of Equation \ref{eq:Lbol} to better account for the likely young ages of these regions. However, \citet{Barnes2021} state that around 20\% of the 5810 regions in their sample have a mass $<10^3 M_{\odot}$, where stochastic effects have a more significant effect on the bolometric luminosity. By using the median for each galaxy, we limit the effect of these potential biases.

In Figure \ref{fig:evolution}, we plot the two pressure components quantified in this work against the radii of the 30 HII regions in this study, along with the values taken from the aforementioned literature sources. Linear fits are plotted for each relationship, with a separate fit for each derivation of the direct radiation pressure. We find that the HII region SMC N76 and the region within Leo P show a significantly higher $P_{\mathrm{dir}}$ than the other regions in this sample. These larger values for $P_{\mathrm{dir}}$ are likely driven by the fact that these observations have a higher spatial resolution ($\lesssim$ 10 pc), and $P_{\mathrm{dir}} \propto r^{-2}$. Therefore, we do not include these regions in the fits for $P_{\mathrm{dir}}$, although further study could make comparisons to observations at similar resolution.

To explore the statistical significance of the relationships investigated in this paper, we perform correlation coefficient analyses using \texttt{pymccorrelation} (\citealt{Curran2014}, \citealt{Privon2020}), whereby we compute the Pearson correlation coefficient, p-value and associated uncertainties for each relationship using a Monte Carlo routine.  With \texttt{pymccorrelation} we perform 1000 bootstrap iterations and 1000 perturbations, according to the `composite method' described in \citet{Curran2014}. The quoted values and their uncertainties for the Pearson correlation coefficients and p-values are the median and the range between the 16th and 84th percentiles of the resulting PDFs.

\subsection{Feedback evolution}
\label{sec:evolution}

As found in other works (e.g. \citealt{Lopez2014}; \citealt{McLeod2019,  McLeod2021};  \citealt{Barnes2021}; \citealt{Olivier_2021}), we see that the pressure of the ionised gas is roughly a few orders of magnitude greater than the direct radiation pressure for HII regions $\sim 10-100$ pc in size, suggesting that photoionisation drives the expansion of these evolved ($T \gtrsim$ 4.5 Myr) HII regions. Similarly to \citet{Olivier_2021} and \citet{Barnes2021}, we find that the pressure terms decrease with HII region size, with a correlation coefficient of $C_{P_{\mathrm{dir,~\texttt{SLUG}}}}=-0.37^{+0.09}_{-0.10}$ and a p-value of $0.003^{+0.030}_{-0.003}$ for $P_{\mathrm{dir,~\texttt{SLUG}}}$, and with $C_{P_{\mathrm{ion}}}=-0.3\pm0.1$ and a p-value of $0.04^{+0.20}_{-0.04}$ for $P_{\mathrm{ion}}$. This indicates only weak negative correlations, and the p-values mean we can only confidently reject the null hypothesis for the direct radiation pressure. We also find that $P_{\mathrm{dir}}$ falls more sharply than $P_{\mathrm{ion}}$ in Figure \ref{fig:evolution}. If we take the size of each region as a proxy for its evolutionary stage, we therefore find that as HII regions evolve and expand, the feedback-driven pressure drops, and the direct radiation pressure becomes less important relative to the ionised gas pressure.

We note that there are caveats with taking HII region size as a proxy for its evolutionary stage. For example, this approach assumes that these HII regions are able to freely expand, and that they have similar stellar populations. There exist other sources of bias in these relationships with $r$;  namely that we have used $r^{-2}$ to calculate our $P_{\mathrm{dir}}$ values, and also that the size of the regions is in part limited by the spatial resolution of the observations, as discussed in Section \ref{sec:hii_regions}. The Milky Way objects studied by \citet{Olivier_2021}, for example, are much smaller and much more compact ($r\lesssim$ 0.5 pc) than the HII regions in this study.

\citet{Olivier_2021} and \citet{Barnes_2020}, who both studied compact HII regions within the Milky Way, also find that the degree of scatter around these pressure-size relationships is more significant for large HII regions, particularly for $P_{\mathrm{dir}}$. \citet{Olivier_2021} find $P_{\mathrm{dir}} \propto r^{-1.36\pm0.16}$, which is significantly steeper than our slopes of $P_{\mathrm{dir, ~IMF}} \propto r^{-0.49(\pm 0.1)}$ and $P_{\mathrm{dir,~\texttt{SLUG}}} \propto r^{-0.66(\pm 0.1)}$. However, our shallower gradients could be a consequence of this increased degree of scatter and of our sampling. Due to the limits of our spatial resolution, we are unable to sample any compact HII regions where these pressure terms have been found to have more importance, and where these relationships have been found to show less scatter. This is particularly true for J0921, at a distance of 21 Mpc and spatial resolution of $\sim$ 83 pc.

An example of the effect of potential biases when using Equation \ref{eq:Lbol} and $P_{\mathrm{dir, ~IMF}}$ to investigate the relationship between the direct radiation pressure and the size of the HII regions is shown in Figure \ref{fig:J0921_P_vs_r} for the HII regions in J0921. Here, we can see that the size of the HII regions is positively correlated to their stellar mass, and therefore for the smaller (and less massive) regions we are likely underestimating $L_{\mathrm{bol}}$ when assuming a fully sampled IMF. In this case, this leads to a strong positive correlation ($C_{P_{\mathrm{dir,~IMF}}} = 0.6^{+0.1}_{-0.2}$ with a p-value of $0.003 ^{+0.04}_{-0.003}$) when using $P_{\mathrm{dir, ~IMF}}$, as opposed to the negative trend described above.

\subsection{Stellar feedback in the low-mass, low-metallicity regime}
\label{sec:low_mass_z}

In Figure \ref{fig:p vs z}, we have plotted the feedback-related pressure components as a function of metallicity, as measured by their oxygen abundance. For this plot, we have also included data from the LMC (\citealt{McLeod2019}), the SMC (\citealt{Lopez2014}, where the oxygen abundance is not measured for each HII region individually but for the SMC as a whole), SMC N76 (McLeod et al. in prep) and from more massive star-forming galaxies from the PHANGS-MUSE survey (\citealt{Barnes2021}). We note that each of the PHANGS-MUSE data points represents the median and standard deviation of multiple HII regions within each of the 19 galaxies in their sample, rather than for an individual HII region as for all the other data points. Since the inclusion of these literature values is intended to show the continuation of the global trends in the pressure terms as a function of metallicity, and since there are $>$ 5000 HII regions in the PHANGS-MUSE catalogue, we choose to only plot these median values for visual clarity. We add that none of the individual PHANGS-MUSE HII regions have an oxygen abundance less than $12+\log(\mathrm{O/H})_{\mathrm{S cal}} \sim 8.1$.

The oxygen abundances of the HII regions in the LMC in \citet{McLeod2019} are determined using the average values given from the N2 and O3N2 indices using equations from \citet{Marino2013}, based on a comparison to values obtained via the direct method. Based on Figure \ref{fig:cdf o abundance}, these values could be as much as 0.3 dex higher than if the S calibration method was used, which would not change the main conclusions of this work. For the SMC, \citet{Lopez2014} assume a metallicity of 20\% solar, which is consistent with the average value obtained from individual regions within the SMC from the direct method (e.g. \citealt{SanCipriano2017}). From Figure \ref{fig:cdf o abundance}, the S calibration method can overestimate the direct method oxygen abundance by as much as 0.6 dex. Again, this would not change our main conclusions. For the PHANGS-MUSE sample, we take the average oxygen abundance for each galaxy given in Table 1 in \citet{Barnes2021}, which are determined using the same calibration (the S calibration of \citealt{Pilyugin2016}) as used for the 30 HII regions in the metal-poor dwarf galaxies studied here.

We also plot these pressure terms for the 30 HII regions studied in this work against their cluster mass in Figure \ref{fig:p vs m}. We do not include literature values in this plot since in most cases they are not quantified, or are derived from different methods that may introduce further biases.

\subsubsection{Direct radiation pressure}
\label{sec:p_dir_discussion}

We find a correlation coefficient of $C=0.65^{+0.07}_{-0.09}$ with a p-value of $(2.75^{+2.73}_{-0.03}) \times 10^{-8}$ for $P_{\mathrm{dir}, ~\texttt{SLUG}}$ against the oxygen abundance for the HII regions in this study and from the literature listed above. This positive correlation is likely driven by two trends: (1) the weak positive correlation we find between $L_{\mathrm{bol}}$ and metallicity ($C=0.3^{+0.1}_{-0.1}$), and (2) the moderate negative correlation we find between the metalllicity and the HII region radius ($C=-0.45^{+0.08}_{-0.08}$). More luminous regions are more metal-rich, and HII regions in metal-poor galaxies tend to be larger. 

The second finding may be expected since HII regions with lower metallicities have higher temperatures (due to less metal-line cooling, e.g. \citealt{Bresolin1999, Kennicutt2000}), and also since low-metallicity stellar populations, at fixed stellar mass, produce more ionising photons (e.g. \citealt{McLeod2021}). These effects increase the thermal pressure, and therefore cause the HII regions to expand faster (\citealt{Ali2021}). Furthermore, the global properties of the host galaxies are also related to the gas-phase metallicity. Metal-poor galaxies tend to be dwarfs with lower ISM pressure and density, which makes it easier for HII regions to expand.

We would also expect the metallicity of HII regions to impact their dust content, and so further impact their size and therefore the direct radiation pressure. One would expect that as the metallicity of a HII region increases, so too does its dust content (\citealt{Remy_Ruyer_2014}), meaning that there is greater extinction of UV photons and the intensity of the ionising spectrum is reduced. A recent study based on radiation-hydrodynamical models of star-forming regions (\citealt{Ali2021}) finds that the presence of dust hinders the expansion of these regions due to this greater extinction of UV photons. That is to say, for a greater dust content, HII regions are smaller, leading to greater values of $P_{\mathrm{dir}}$. This predicted positive correlation between dust and $P_{\mathrm{dir}}$ is supported by observations of the HII regions within the LMC, SMC and NGC 300 (\citealt{Lopez2014}; \citealt{Roman_Duval_2014}; \citealt{McLeod2021}). However, for the 30 HII regions in this study we find no correlation  for size ($C=-0.0\pm0.2$)  nor $P_{\mathrm{dir}}$ ($C=-0.1^{+0.1}_{-0.3}$, i..e, consistent with zero) as a function of the dust content, as measured by the visual extinction, $A_V$ (assuming that the extinction, measured from the Balmer decrement, is mainly originating from dust within the galaxies rather than foreground material). It is therefore unclear whether a change in the dust content is driving these trends, although we note that the unknown extinction curve for these galaxies is a source of uncertainty when determining the extinction. However, a test with the \citet{Cardelli1989} Galactic extinction curve produces similar findings to those presented here.

Due to the high uncertainty in the derived mass values, the trends observed with the cluster mass show a high degree of scatter. However, as with Figure \ref{fig:J0921_P_vs_r}, we see that different conclusions can be drawn from HII region studies if stochastic effects are not taken into account. A moderate correlation is found between $P_{\mathrm{dir,~ IMF}}$ and the cluster mass with $C=0.4\pm0.2$. However, when $P_{\mathrm{dir,~\texttt{SLUG}}}$ is used, $C$ is consistent with zero. This is again likely driven by the consequences of using Equation \ref{eq:Lbol} to estimate $L_{\mathrm{bol}}$. $L_{\mathrm{bol}}$ of the lower mass regions is underestimated, and likely overestimated for the higher mass and younger regions, particularly in KKH046 (Figure \ref{fig:p dir comparison}).

Overall, we tentatively see that HII regions in low-mass, low-metallicity host galaxies tend to have a larger radius and hence a lower value of $P_{\mathrm{dir}}$. A larger sample and higher resolution observations would be needed to investigate these trends fully.

\subsubsection{Pressure of the ionised gas}
\label{sec:pion_discussion}

The trend found for $P_{\mathrm{ion}}$ as a function of metallicity in Figure \ref{fig:p vs z}, is found to be much weaker than for $P_{\mathrm{dir}}$, though we see a slight negative correlation with $C=-0.14^{+0.24}_{-0.09}$ and a p-value of $0.35^{+40}_{-0.29}$. Whilst we cannot reject the null hypothesis, a negative correlation between photoionisation pressure and metallicity follows from the expectation that metal-poor HII regions should have a higher temperature (e.g. \citealt{Kennicutt2000}), and then $P_{\mathrm{ion}} \propto T_e$. In terms of investigating $P_{\mathrm{ion}}$ as a function of metallicity, and other quantities, we are limited by assuming an electron temperature of $10^4$ K for all HII regions. To further investigate these relationships, we would benefit from deeper spectra that would allow us to determine the electron temperature, and also use the direct method for determining the oxygen abundance.

For $P_{\mathrm{ion}}$ as a function of cluster mass in Figure \ref{fig:p vs m}, we find a weak negative correlation with $C=-0.2\pm0.1$ and a p-value of $0.2^{+0.4}_{-0.2}$. Based on the p-value and the large uncertainties in the cluster mass, we again cannot reject the null hypothesis.  However, in Figure \ref{fig:p vs m}, we see that the more massive regions tend to have a larger radius. Following from the trend in Figure \ref{fig:p vs r}, this may then drive a decrease in $P_{\mathrm{ion}}$ with increasing mass.

We note that the relationships between $P_{\mathrm{dir}}$ and $P_{\mathrm{ion}}$ also persist when only the regions with direct method metallicities are considered (see Appendix \ref{appendix:oxygen abundances}). However, a more statistically significant sample with direct method metallicities is necessary to investigate these trends further.

\section{Summary \& conclusions}
\label{sec:conc}

In this paper, we have presented the results of a stellar feedback study of HII regions within three nearby, low-metallicity, dwarf starburst galaxies (J0921, KKH046, and Leo P), which were observed with the IFU spectrograph, MUSE, and collected by the DWALIN collaboration. We have discussed the methods used to select and extract the integrated spectra of these HII regions via the \texttt{ASTRODENDRO} Python package, and the methods used to derive their key ionised gas properties. We have also presented the use of stochastically sampled stellar population synthesis models using the \texttt{SLUG} package to investigate the properties of the ionising stellar populations within these HII regions. Finally, we have quantified two pre-SN stellar feedback mechanisms and quantified them as a function of the selected environmental properties. This has allowed us to investigate the evolution of stellar feedback within low-mass, low-metallicity HII regions, and make comparisons to observations of more massive, metal-rich star-forming galaxies from the recently published results of the PHANGS-MUSE survey (\citealt{Barnes2021}).

The key findings of this paper are summarised as follows:
\begin{enumerate}
   
    \item There are considerable discrepancies between the different methods for determining the oxygen abundance using strong-line ratios. To increase our confidence in the oxygen abundance values we have reported, and therefore in our interpretations of the results, deeper spectra of these galaxies covering temperature-sensitive auroral lines for the direct method would be necessary.
    
    \item We estimate that the HII regions within our sample are low-mass ($M_* \lesssim 6 \times 10^4 M_{\odot}$) and evolved ($T \gtrsim 4.5$ Myr) regions and we therefore believe that their bolometric luminosities are underestimated in most cases if one assumes a fully sampled IMF. 
    \item We see that stochastic sampling effects can significantly affect the trends observed between feedback-related pressure terms and properties of the HII regions. This stresses the importance of stochastic sampling in the low-mass regime.
    \item Our simulated \texttt{SLUG} libraries for Leo P are consistent with a previously reported late O-type star within Leo P's HII region, which we find to be of mass $23^{+4}_{-7} M_{\odot}$ and age $8^{+2}_{-1}$ Myr.  This illustrates the success of \texttt{SLUG} at identifying massive stars from observations that may not have the necessary spatial resolution for traditional techniques.
    \item For all HII regions in our sample, the pressure of the ionised gas is greater than the direct radiation pressure. This supports the findings of feedback studies of other giant HII regions, such as in \citet{Olivier_2021} and \citet{Barnes2021}. This suggests that photoionisation is more dominant than the direct radiation pressure within these evolved HII regions.
    \item The two feedback mechanisms that we have quantified are shown, overall, to decrease with region size, supporting literature findings that these feedback mechanisms have less of an importance in giant, evolved HII regions when compared to ultra-compact HII regions.
    \item We find that the direct radiation pressure within HII regions increases with metallicity. In some cases,  this may follow from more rapid expansion in metal-poor, and hence hotter regions.
    \item Deeper spectra covering the auroral lines necessary for determining the electron temperature and the oxygen abundance via the direct method are required to fully investigate relationships between $P_{\mathrm{ion}}$ and the ionised gas and stellar properties of HII regions.
\end{enumerate}

In conclusion, we find that an investigation of stellar feedback in the extreme environments of low-metallicity, dwarf starburst galaxies, along with comparisons to more massive systems, can provide a wide range of physical properties to probe the environmental dependence of stellar feedback mechanisms. Follow-up investigations, that is, deeper spectroscopic observations, IR, and X-ray data, and investigations into the full sample of star-forming galaxies provided by DWALIN and other upcoming surveys will provide further observational constraints on these mechanisms of stellar feedback and how they affect the evolution of galaxies.



\begin{acknowledgements}
The authors thank the anonymous referee for their constructive comments that have improved the clarity and quality of the paper. The authors also thank Eve Cully for work on the analysis of the HII region SMC N76.

This paper made use of the following software
packages: Astropy (\citealt{astropy2013}; \citealt{astropy2018}), Matplotlib (\citealt{Hunter2007}), NumPy (\citealt{harris2020array}), SciPy (\citealt{2020SciPy-NMeth}), spectral-cube (\citealt{SpectralCube}), and astrodendro (http://www.dendrograms.org/). AF is supported by a UKRI Future Leaders Fellowship (grant no. MR/T042362/1). FB acknowledges support from the INAF Ricerca Fondamentale 2022 Grant. MRK acknowledges funding from the Australian Research Council through its Laureate Fellowship scheme, award FL220100020. GV acknowledges support from ANID programme FONDECYT Postdoctorado 3200802.

\end{acknowledgements}

\bibliographystyle{aa} 
\bibliography{library}

\begin{appendix}

\section{J0921 - Comparison with previous studies}
\label{appendix:j0921 comparison}

\begin{figure*}[h]
    \centering
    \begin{subfigure}[b]{0.45\textwidth}
        \centering
        \includegraphics[width=\textwidth]{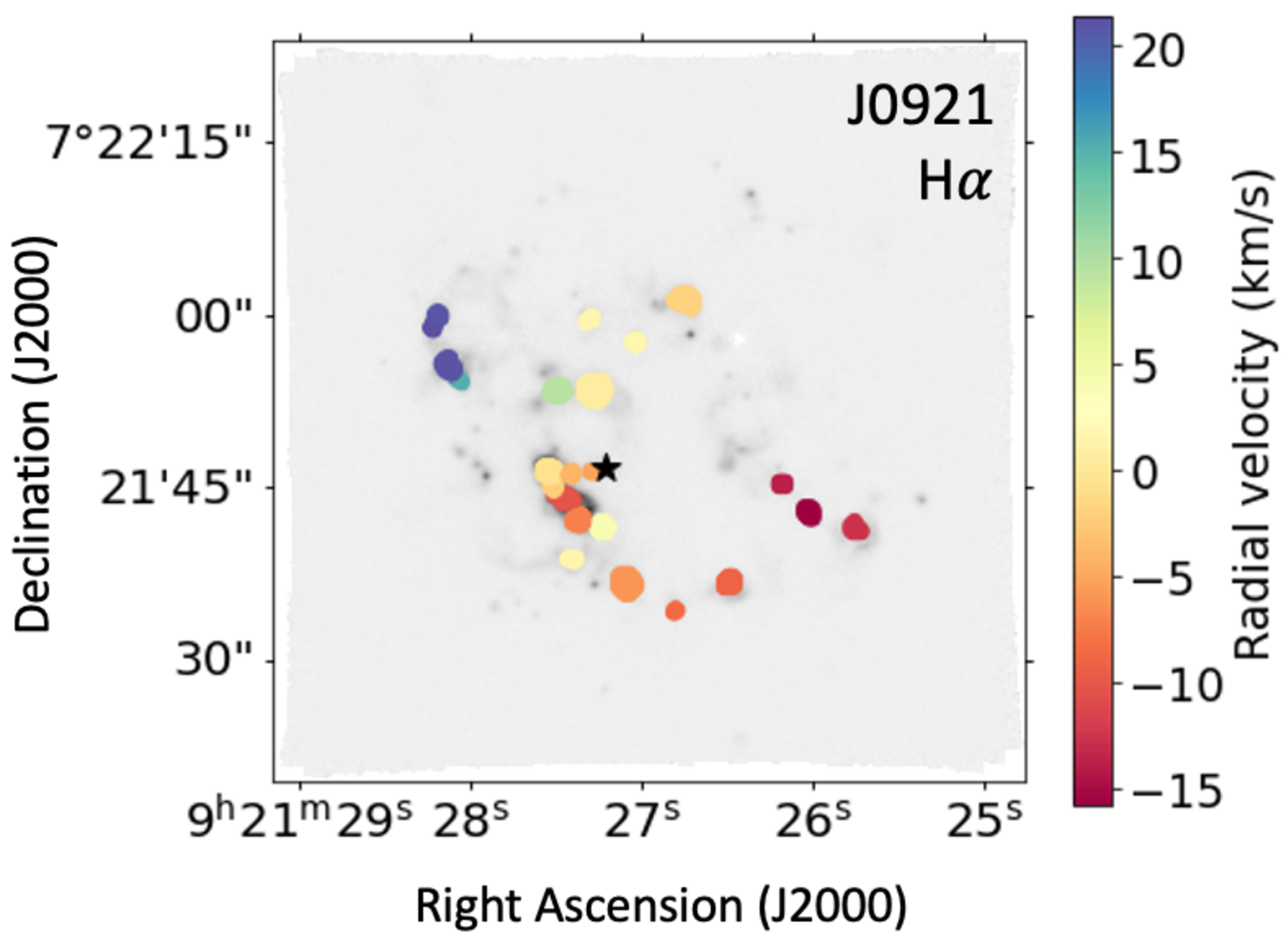}
        \caption{}
        \label{fig:J0921 velocity}
    \end{subfigure}
    \begin{subfigure}[b]{0.45\textwidth}
        \centering
        \includegraphics[width=\textwidth]{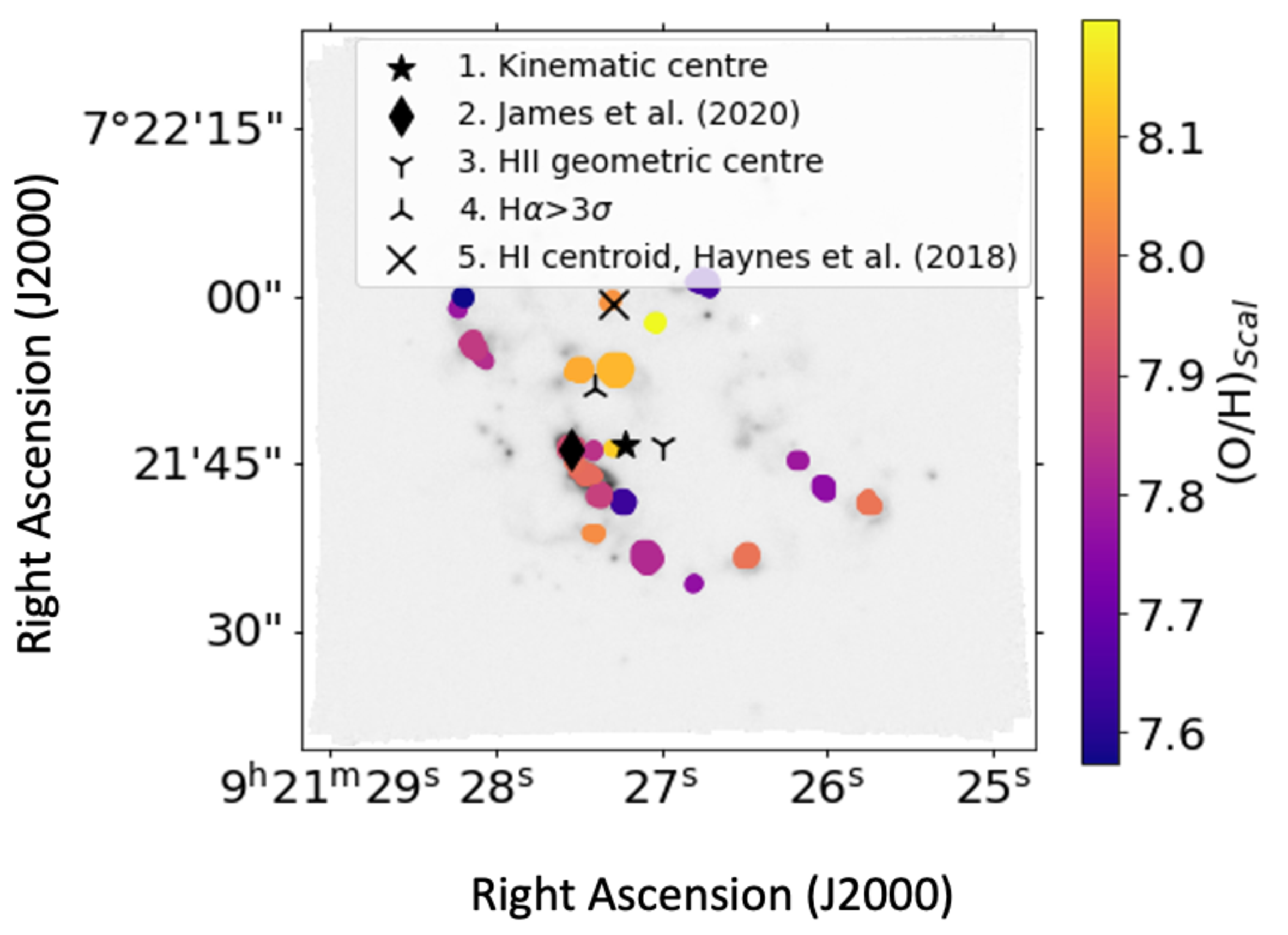}
        \caption{}
        \label{fig:J0921 o abundance map}
    \end{subfigure}
    
   \textbf{Figure A1:} Panel (a) shows the H$\alpha$ emission map of J0921, with each region coloured according to its radial velocity (not inclination-corrected due to the unknown inclination of the system). Red indicates regions moving away from us, and blue indicates regions moving towards us. Here we can see evidence that J0921 is rotating as a large body, as discussed in \citetalias{J20} and also illustrated in \citet{Marasco2022}. Panel (b): Similar to (a), but regions are instead coloured by their oxygen abundance (using the S calibration) to indicate the spatial distribution of metals in J0921. Also plotted are the various galactic centres tested in this work, as listed in the text.
    \vspace{-1em}

\end{figure*}

\begin{figure*}[h]
    \centering
    \begin{subfigure}[b]{0.45\textwidth}
        \centering
        \includegraphics[width=0.7\textwidth,angle=270]{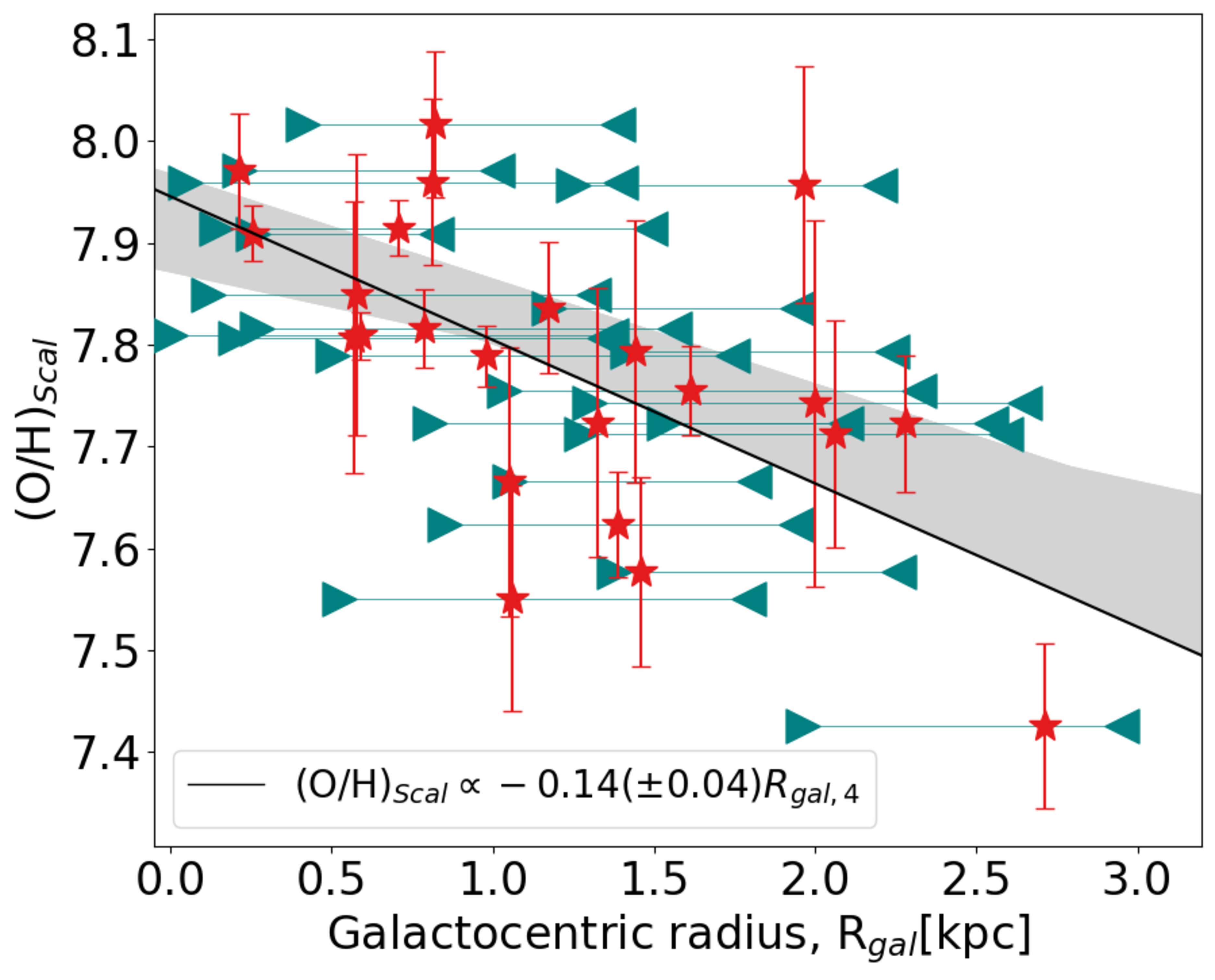}
        \caption{}
        \label{fig:Z gradient}
    \end{subfigure}
    \begin{subfigure}[b]{0.45\textwidth}
        \centering
        \includegraphics[width=0.7\textwidth,angle=270]{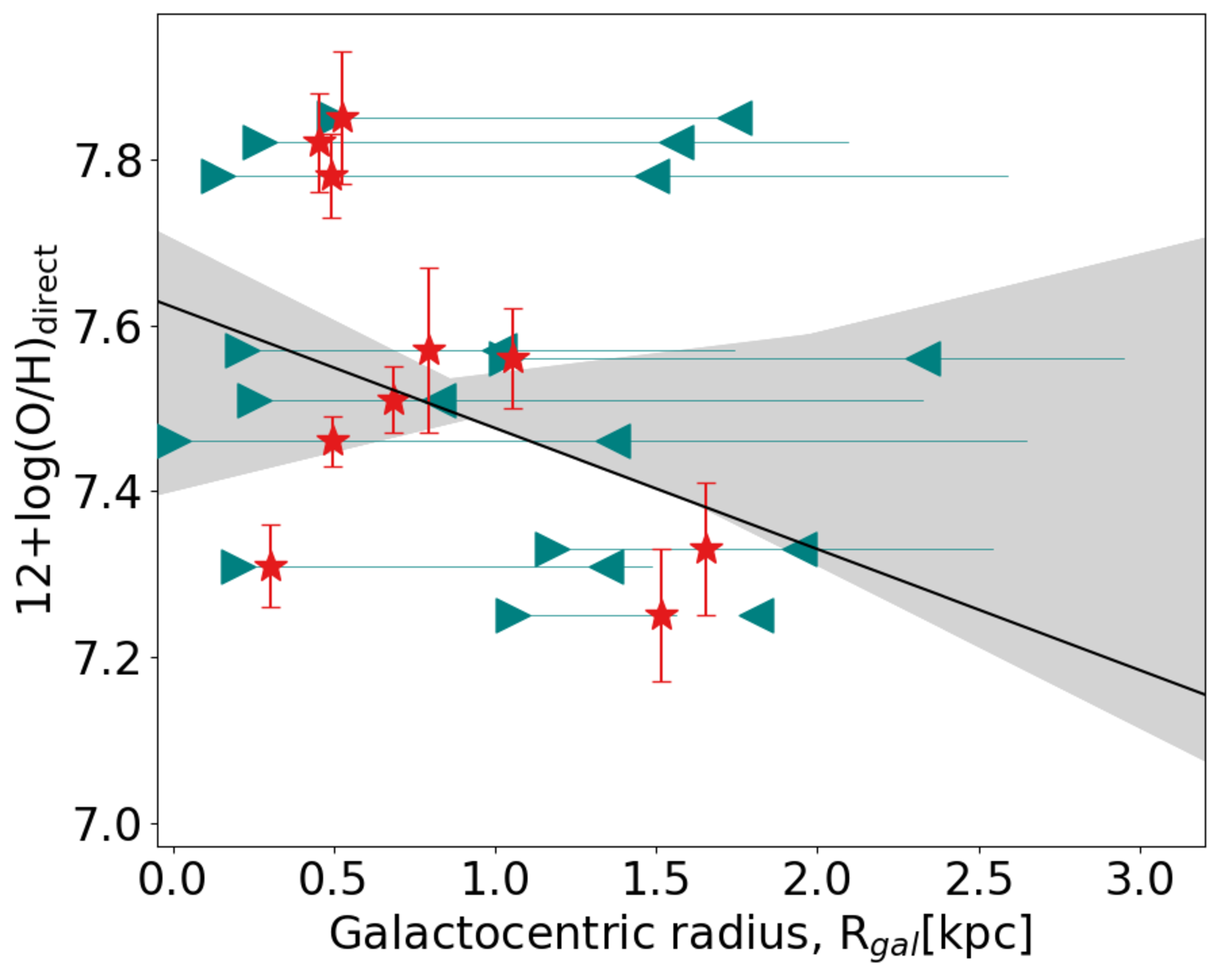}
        \caption{}
        \label{fig:bootstrap}
    \end{subfigure}
    
    \textbf{Figure A2:} Panel (a): Oxygen abundance against projected galactocentric distance for the 23 HII regions within J0921. Horizontal teal lines indicate the change in the position of each region when different galactic centres were used, and the shaded grey area shows the spread of metallicity gradients for the five galactic centres tested. The points plotted in red are for the fourth centre, which is the geometric centre of all pixels containing H$\alpha$ emission above three standard deviations. Panel (b): Similar to panel (a) but with the direct method oxygen abundance for the ten regions from \citetalias{J20}.

\end{figure*}

Here, we focus on some of the ionised gas properties derived for one of our target dwarf galaxies, J0921, since it contains $\sim 77\%$ of the HII regions in our sample and has also been the focus of a previous study, \citetalias{J20}. \citetalias{J20} used the same MUSE data to explore the chemical homogeneity of J0921, referred to as JKB 18 in their study (from the JKB catalogue presented in \citealt{James2015} and \citealt{James2017}).

In \citetalias{J20}, an IFAUNAL binning algorithm with an optimum bin size of seven pixels and a minimum flux threshold of $10^{-19} \mathrm{erg s^{-1}}$ is used to select HII regions (i.e. a lower flux threshold then in our methods described in Section \ref{sec:hii_regions}). This resulted in 30 HII regions. All 23 HII regions identified in this study have a match within \citetalias{J20} (note region 14 in \citetalias{J20} encompasses both regions 20 and 22 in this study) and the derived values of $L_{H\alpha}$ and $n_e$ between \citetalias{J20} and this study are comparable (within the uncertainties). Oxygen abundance values were determined using the O3N2 index in \citetalias{J20}, however in this study we instead use the S calibration method.

We carried out similar tests as in \citetalias{J20} to search for evidence of a metallicity gradient within this galaxy. In \citetalias{J20}, the centre of the galaxy is taken to be the centre of the brightest HII region. We investigated additional galactic centres, with 5 centres tested in total: (1) an estimated kinematic centre (Figure A1a), (2) the centre of the brightest region, (3) the geometric centre of all HII regions, (4) the geometric centre of all pixels with H$\alpha$ emission above a 3$\sigma$ detection and (5) the HI centroid from \citet{Haynes2018}. The position of these centres are labelled in Figure A1b. For all of these potential galactic centres, a slight negative metallicity gradient is found when the S calibration method is used (Figure A2a; note that the galactocentric radii, $R_{gal}$, values are projected on the plane of the sky due to the unknown inclination of the system). However, when the direct method metallicity values for the subsample of 11 HII regions with a $>3 \sigma$ detection of the auroral [SII]$_{\lambda 6312}$ line (given by \citetalias{J20}) are used with these galactic centres, no convincing gradient is found (Figure A2b). Similarly to \citetalias{J20}, we therefore conclude that there is insufficient evidence to support the presence of a negative metallicity gradient within this galaxy. The weak correlation found when using the strong line methods is perhaps driven by some other radial change within this galaxy, such as a change in the ionisation parameter. 

In \citet{monreal_ibero_2023}, a similar discrepancy is found when studying the spatial distribution of metals within the nearby dwarf galaxy UM 462. When using the direct method oxygen abundance, the galaxy appears chemically homogeneous. However, when using strong-line methods, chemical inhomogeneities were identified. We therefore support  \citet{monreal_ibero_2023} in stating that further work is necessary to determine the optimal method for calculating the oxygen abundance at any point within a galaxy.

\section{Radial intensity profiles}
\label{appendix:radial intensity profiles}
\begin{figure}[h!]
    \centering
    \includegraphics[width=0.5\textwidth]{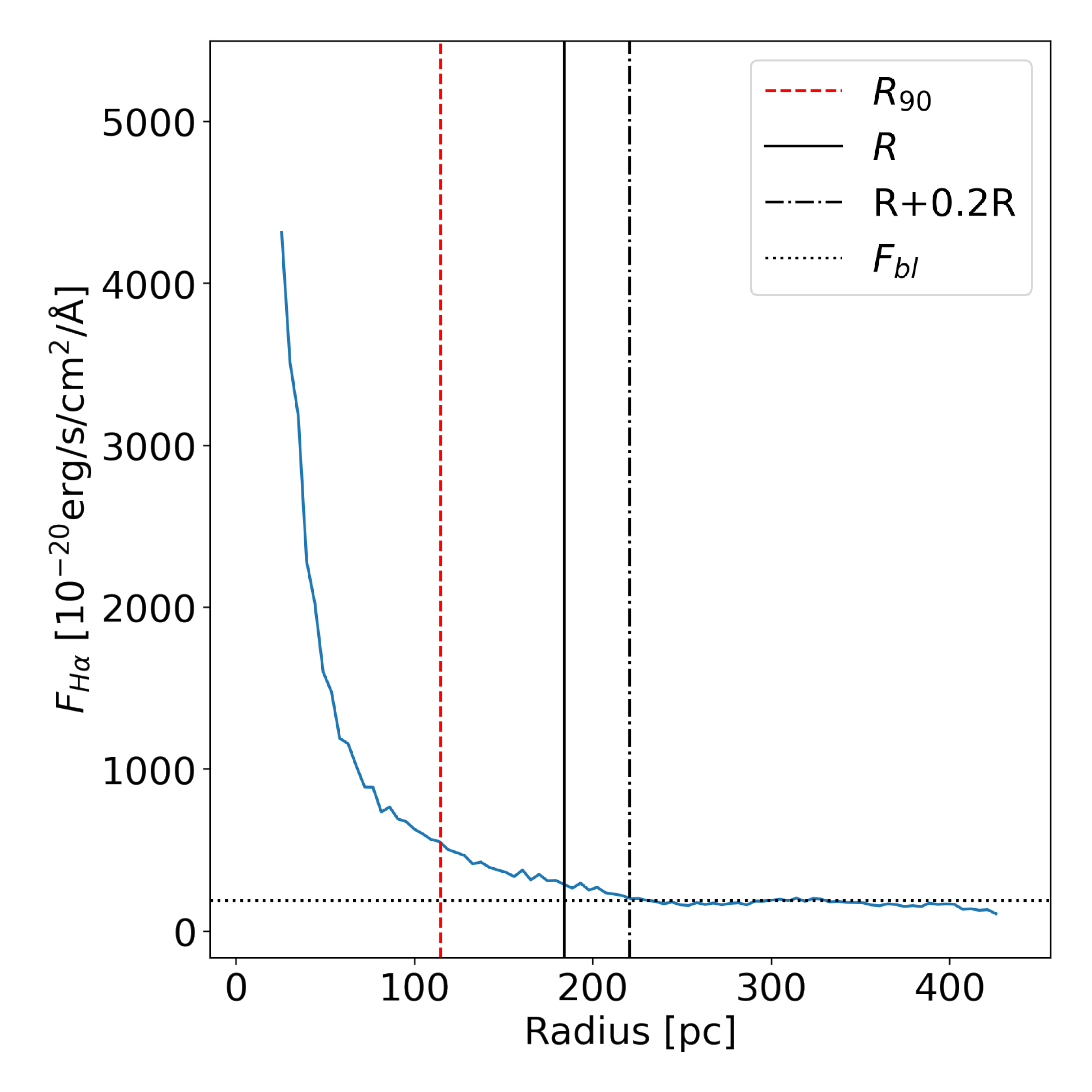}
    \textbf{Figure B1:} Radial flux profile for KKH046 region 1. The dashed red line is the radius that encompasses 90\% of the H$\alpha$ emission and the solid black line is the radius returned by \texttt{ASTRODENDRO}. The horizontal line depicts where the flux profile reaches the background level, $F_{bl}$ (median of the emission). This line was used to determine the uncertainty in the size of these regions, which we determined to be 20\%.
    \label{fig:fig1A1}
\end{figure}

To test the sizes of the HII regions studied here, radial flux profiles for a box (of size judged by eye) surrounding each visibly distinguishable ionised gas clump in the H$\alpha$ emission maps were plotted. The radius that encompasses 90\% of the H$\alpha$ flux in each box, $R_{90}$, was then calculated and compared to the radius of the region returned by \texttt{ASTRODENDRO}. An example radial flux profile for region KKH046 1 is shown in Figure B1. The $R_{90}$ values for all regions across all three galaxies were found to be comparable to the radius of our final dendrogram structures (measured by the geometric mean of the major and minor axes), which are listed in column 2 of Table \ref{tab:results}.  However, it must be noted that \texttt{ASTRODENDRO} calculates the radius using the geometric mean of the major and minor axes, whereas the $R_{90}$ calculation assumes the regions are circular.

\section{Uncertainty estimation}
\label{appendix:uncertainties}

The uncertainties in this work consist of those returned by \texttt{PYSPECKIT} for the Gaussian-fit parameters of each emission line, the uncertainties in each of the calibrations we use based on emission lines, the 20\% uncertainty on the radius, $r$, of our regions (as estimated by the radius at which the radial intensity profiles reach the background level, see Figure B1), the uncertainty in the extinction correction, and the 68\% confidence intervals of our posterior PDFs for the cluster mass, age and bolometric luminosity. We note that there is an additional uncertainty in the true extinction curves of these galaxies, which could in fact be different from that given by \citet{Calzetti2000}, and additional uncertainties in the calibrations using emission line ratios. In particular, studies have found that stochasticity may have an effect on all of these calibrations, for example readers can refer to \citet{D_Agostino_2019}. We also note that different papers often use different methods and calibrations, and there will therefore exist some discrepancies between values. This must be taken into account when making comparisons between different literature values.

\section{Oxygen abundance calibration tests}
\label{appendix:oxygen abundances}

In Section \ref{sec:oxygen} we derived the oxygen abundances of our HII regions using the S calibration method of \citet{Pilyugin2016}. In Section \ref{sec:oxygen}, we discussed some of the discrepancies between different calibrations. Here, we continue this discussion.

We investigated the effect of the calibration method when quantifying the feedback-related pressure terms as a function of metallicity. For the overall trends including literature values, the different calibrations have little effect and are similar to those in Figure \ref{fig:p vs z}. However, when only considering the 30 HII regions in this study, the N2 and O3N2 abundances show a slight negative gradient for $P_{\mathrm{dir}}$ against 12+log(O/H), whereas the S cal, \citet{Dopita} and direct method abundances show a slight positive gradient. We show the trends when using the available direct method abundances and when using the N2 index in Figures D1 and D2, respectively. We expect that the N2 and O3N2 indices are not a good measure of the oxygen abundance for these regions due to their dependence on the ionisation parameter, as discussed in \citet{Kewley2019b}.

\begin{figure}[h]
    \centering
    \includegraphics[width=0.45\textwidth,angle=270]{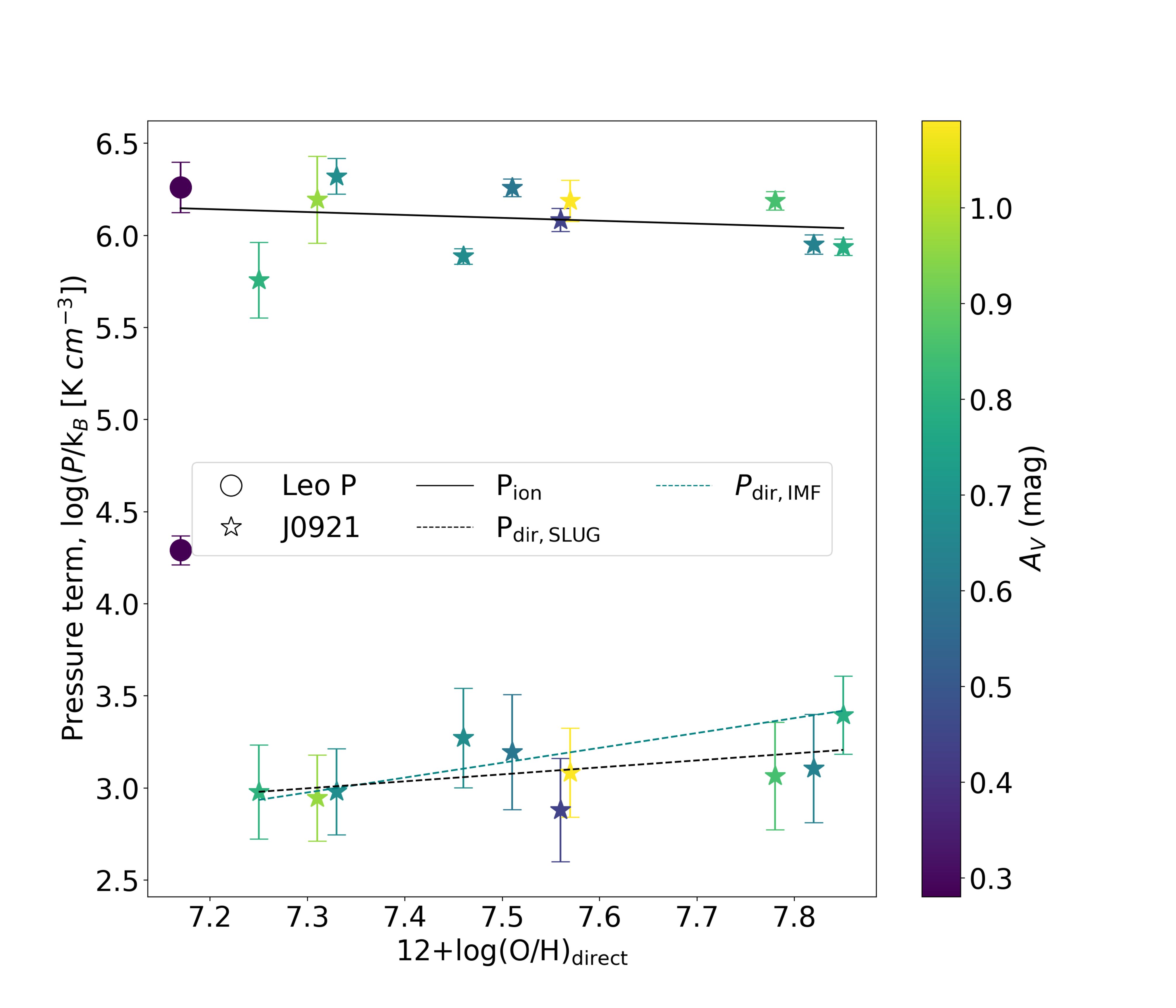}
    
    \textbf{Figure D1:} Similar to Figure \ref{fig:p vs z}, but with the available direct method oxygen abundances from \citetalias{J20} and \citet{McQuinn_2015} (11 HII regions within this study).
    \label{fig:figC1}
\end{figure}

\begin{figure}[h]
    \centering
    \includegraphics[width=0.4\textwidth,angle=270]{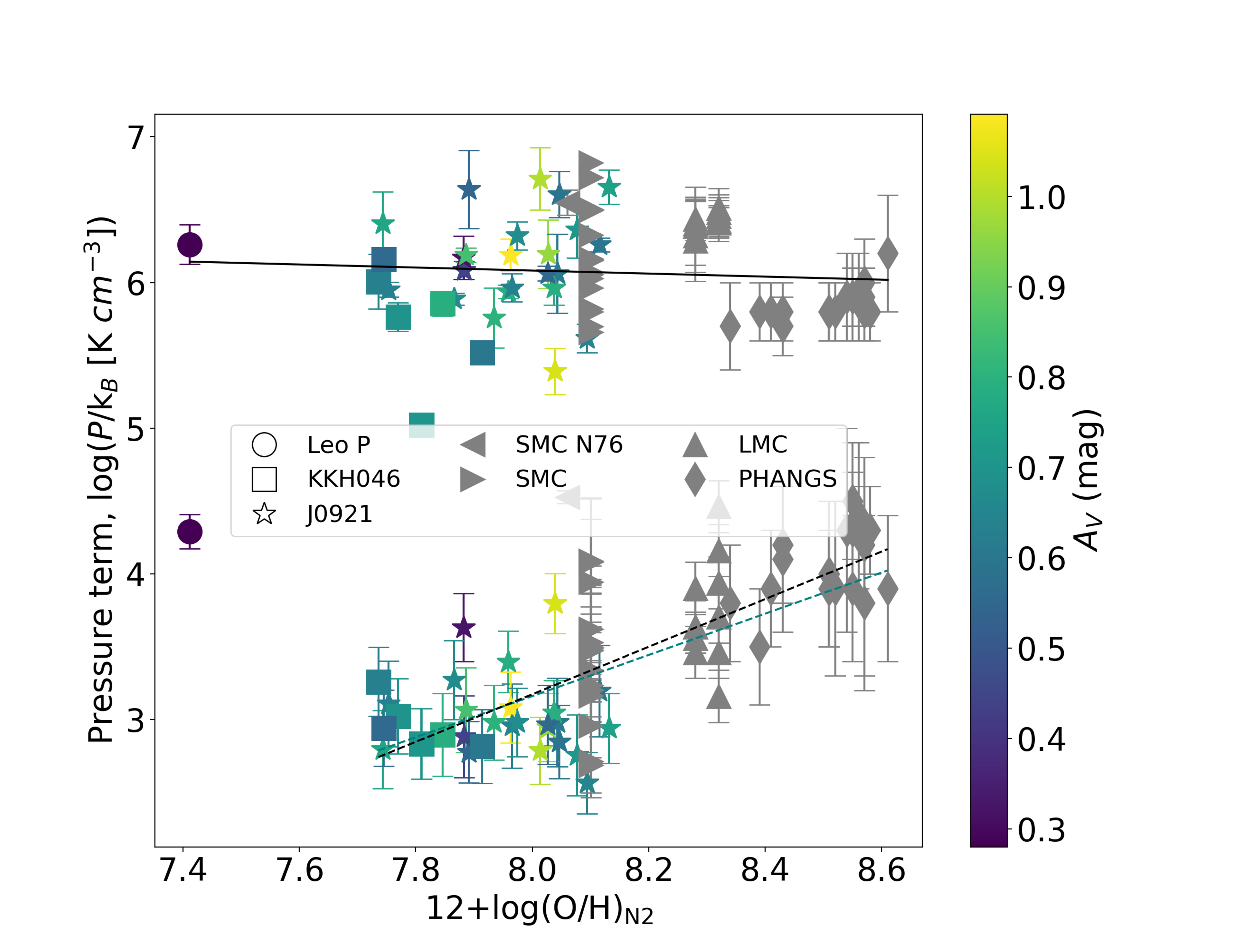}
    
    \textbf{Figure D2:} Similar to Figure \ref{fig:p vs z}, but with the pressure components instead plotted against the oxygen abundance derived using the N2 index.
    \label{fig:figC2}
\end{figure}

\end{appendix}

\end{document}